\documentclass[aps,twocolumn,prl,floatfix,groupedaddress,nofootinbib,notitlepage,showpacs,superscriptaddress]{revtex4-1}

\usepackage{amsmath}
\usepackage{amsthm}
\usepackage{latexsym}
\usepackage{amssymb}
\usepackage{enumitem}
\usepackage{bm}
\usepackage{bbm}
\usepackage{appendix}
\usepackage{graphics,epstopdf}
\usepackage{physics}
\usepackage{color}
\usepackage[colorlinks=true,linkcolor=blue,citecolor=green,plainpages=false,pdfpagelabels]{hyperref}
\usepackage[normalem]{ulem}
\usepackage{newlfont}
\usepackage{amsfonts}
\usepackage{amsthm}
\usepackage{times}
\usepackage{braket}
\usepackage{graphicx}
\usepackage{epsfig}
\usepackage{mathrsfs}
\usepackage[thinc]{esdiff}
\usepackage{caption}
\usepackage{subcaption}

\usepackage{times}

\DeclareOldFontCommand{\rm}{\normalfont\rmfamily}{\mathrm}

\usepackage{bigints}

\usepackage{amsthm}
\usepackage{latexsym}
\usepackage{float}
\usepackage{appendix}
\usepackage{epstopdf}
\usepackage{physics}
\usepackage[colorlinks=true,linkcolor=blue,citecolor=green,plainpages=false,pdfpagelabels]{hyperref}
\usepackage[normalem]{ulem}
\usepackage{newlfont}
\usepackage{amsfonts}
\usepackage{epsfig}
\usepackage{mathrsfs}
\usepackage[thinc]{esdiff}

\DeclareOldFontCommand{\rm}{\normalfont\rmfamily}{\mathrm}
\usepackage{bigints}
\usepackage{braket}
\usepackage{caption}
\usepackage{subcaption}
\usepackage{setspace}
\usepackage{amsmath}
\usepackage{times,bm,bbm,bbold,graphicx,graphics,amssymb,amsmath,amsfonts,dsfont,hyperref,color}
\usepackage{mathrsfs,cancel}
\usepackage[utf8]{inputenc}
\usepackage{soul,xcolor}
\setstcolor{red}
\definecolor{mycolor}{rgb}{0.1, 0.1, 0.7}
\usepackage{dsfont}
\usepackage{url}
\usepackage{soul}
\usepackage[vlined,ruled]{algorithm2e}
\DeclareFontFamily{OT1}{pzc}{}
\DeclareFontShape{OT1}{pzc}{m}{it}%
{<-> s * [1.25] pzcmi7t}{}
\DeclareMathAlphabet{\mathpzc}{OT1}{pzc}%
{m}{it}
\hypersetup{
	plainpages=true,
	breaklinks=true,
	hypertexnames=false,
	pageanchor=true,
	colorlinks=true,
	linkcolor={mycolor},
	citecolor={mycolor},
	urlcolor={mycolor},
	anchorcolor={black}
}

\begin{document}

\title{Unified Framework for Direct Characterization of Kraus Operators, Observables, Density Matrices, and Weak Values Without Weak Interaction}

\author{Sahil}
\email{sahil402b2@gmail.com}
\affiliation{Optics and Quantum Information Group, The Institute of Mathematical Sciences, CIT Campus, Taramani, Chennai 600 113, India}
\affiliation{Homi Bhabha National Institute, Training School Complex, Anushakti Nagar, Mumbai 400 085, India}

\author{Sohail}
\email{sohail.sohail@ttu.edu}
\affiliation{Department of Computer Science, Texas Tech University, Lubbock, TX 79409, USA}

\begin{abstract}
Generalized quantum measurements, described by positive operator-valued measures (POVMs), are essential for modeling realistic processes in open quantum systems. While quantum process tomography can fully characterize a POVM, it is resource-intensive and impractical when only specific POVM elements or matrix elements of a particular POVM element are of interest. Direct quantum measurement tomography offers a more efficient alternative but typically relies on weak interactions and complex structures of the system, environment, and probe as the dimension of the system increases, limiting its precision and scalability. Furthermore, characterizing a POVM element alone is insufficient to determine the underlying physical mechanism, as multiple Kraus operators can yield the same measurement statistics. In this work, we present a unified framework for the direct characterization of individual matrix elements of Kraus operators associated with specific POVM elements and arbitrary input states—without requiring weak interaction, complex structures of the system-environment-probe or full process and state tomography. This framework naturally extends to projective measurements, enabling direct observable tomography, and to the characterization of unitary operations. Our method also captures modular and weak values of observables and Kraus operators, without invoking weak interaction approximations. We demonstrate potential implementations in optical systems, highlighting the experimental feasibility of our approach.
\end{abstract}

\maketitle

\emph{Introduction.---} Positive operator-valued measures (POVMs) lie at the heart of quantum measurement theory, providing a comprehensive framework for describing generalized quantum measurements beyond ideal projective scenarios \cite{Nielsen-Chuang-2010,Jacobs-2014}. In practical settings—particularly where quantum systems interact continuously with an environment—POVMs capture the full range of measurement statistics accessible to observers. A conventional approach to characterizing such measurements is quantum process tomography, which reconstructs the entire set of POVM elements \cite{Nielsen-Chuang-2010,Mohseni-Rezakhani-Lidar-2008}. While robust, this method is resource-intensive and inefficient when only partial information about a specific POVM element—or a few of its matrix elements—is required, since it demands reconstruction of the entire measurement process.\par
To address this limitation, recent work has developed \textit{direct characterization of quantum measurements} (DCQM), which enables the selective determination of individual matrix elements of a POVM element with substantially fewer resources~\cite{Xu-2021-DirectMeasurements, Kim-Kim-Lee-Han-Cho-2018}. These protocols, however, typically rely on weak or strong system--probe interactions and often require ancillary degrees of freedom or carefully engineered environments. Consequently, their precision is limited and the experimental complexity grows unfavorably with system dimension.\par
Beyond reconstructing POVM elements, a more complete understanding of measurement processes requires identifying the specific Kraus operators responsible for generating the observed statistics. This task is nontrivial: a single POVM element can correspond to multiple, physically distinct Kraus operators. For example, consider the POVM \(\left\{ \frac{1}{2}I, \frac{1}{8}(I + \sigma_Z), \frac{1}{8}(3I - \sigma_Z) \right\}\), where \(\sigma_Z\) is the Pauli \(Z\) operator. The element \(E_0 = \frac{1}{2}I\) admits multiple valid Kraus representations, such as \(A_0 = \frac{1}{\sqrt{2}}I\) and \(\widetilde{A}_0 = \frac{1}{2}(\sigma_X + \sigma_Z)\), with \(\sigma_X\) the Pauli \(X\) operator, both satisfying \(E_0 = A_0^\dagger A_0 = \widetilde{A}_0^\dagger \widetilde{A}_0\). This ambiguity highlights the need for methods that identify not just POVM elements, but also the Kraus operators underlying their physical implementation.\par
Parallel to DCQM, accurate characterization of the quantum state—particularly the off-diagonal elements of the density matrix—is essential for identifying non-classical features such as coherence and entanglement, which underpin many emerging quantum technologies \cite{Horodecki2009,Modi2010,Streltsov2017,Degen2017,Braun2018,Giovannetti2011,Vidrighin2014}. The standard approach to obtaining complete information about a quantum state is quantum state tomography (QST), which reconstructs the full density matrix using an informationally complete set of measurements \cite{James-2001,Thew-2002}. However, both the experimental and computational requirements grow rapidly with system size, which makes the QST technique unsuitable for large-scale systems.\par
To circumvent these challenges, direct characterization of the density matrix (DCDM) has been introduced as an efficient alternative for selectively measuring individual matrix elements \cite{Lundeen‑Sutherland‑Patel‑Stewart‑Bamber‑2011,Lundeen-Bamber-2012,Thekkadath-2016,Vallone-Dequal-2016,Vallone-2018,Xu-Zhou-2024}. Rather than reconstructing the entire state, DCDM focuses on estimating specific entries, thereby reducing resource requirements. Despite this advantage, current implementations of DCDM typically rely on weak measurement techniques and sequential coupling schemes, which limits both efficiency and scalability in high-dimensional or multipartite settings. Moreover, extracting high-order correlation functions via sequential weak measurements tends to amplify statistical errors, thereby compromising accuracy \cite{Lundeen-Bamber-2012,Thekkadath-2016}. To improve the precision of DCDM, frameworks based on strong measurements have been developed \cite{Vallone-Dequal-2016,Vallone-2018}, offering enhanced accuracy. However, these approaches may compromise the key advantage of weak measurements by inducing stronger measurements to the quantum system and often requires additional assumptions or auxiliary resources.\par
A common feature of both DCQM and DCDM methods is the need to implement a set of  $d$  unitary evolution operators, typically constructed from system projection operators forming a complete measurement basis in order to fully characterize the unknown POVM element or density matrix of the system. Moreover, if one requires only single or very few matrix elements of a POVM element or density matrix, then a large amount of resource will be wasted in the DCQM or DCDM methods.  This is because all elements of a complete measurement basis will inevitably be occurred, regardless of how many projection operators out of complete measurement basis is actually required. Additionally, existing DCQM and DCDM protocols have been developed under disparate assumptions, including variations in measurement strength, number of required operations, and experimental architecture. As such, there remains a lack of a unified framework capable of simultaneously supporting both DCQM and DCDM in a scalable and resource-efficient manner.\par
In this work, we present a unified and versatile framework for the direct characterization of Kraus operators associated with specific POVM elements, as well as unknown quantum states, without relying on complex system-environment-probe interactions or full quantum process and state tomography. In the special case where the Kraus operators reduce to projectors, the corresponding measurement describes a quantum observable. Using this connection, we introduce a method for direct characterization of an unknown observable, enabling the reconstruction of observables through a procedure that is conceptually very closely related—but not identical—to our Kraus operator characterization scheme. Our framework further extends naturally to the direct estimation of unitary operators of quantum systems. It also includes derivations of modular values and weak values of observables and Kraus operators, which serve as foundational tools for direct characterization of observables. Crucially, all results presented here are obtained without invoking the weak coupling approximation; our approach operates effectively in both weak and strong coupling regimes, thereby eliminating a common constraint in existing methods.\par
Notably, our framework simplifies experimental implementation by requiring only a single unitary evolution operator (\emph{e.g.,} a $d$-dimensional Hadamard gate) for Kraus operator characterization, and approximately $d/2$ Pauli X-gates for density matrix characterization. This contrasts with the $d$ unitary operations required in conventional DCQM and DCDM, offering a significant reduction in complexity and enhancing feasibility in high-dimensional systems.
\vspace{2mm} 
\\
\emph{Unified framework.}---To define Kraus operators within the framework of quantum measurement theory, an interaction between the system and its environment is typically required. In our approach, we introduce a probe that interacts with both the system and the environment, enabling the direct characterization of individual Kraus operators. For characterizing the density matrix alone, however, the system-environment interaction can be neglected, as it plays no essential role. Let $\mathcal{H}_{\text{P}}$, $\mathcal{H}_{\text{S}}$, and $\mathcal{H}_{\text{E}}$ denote the Hilbert spaces of the probe, system, and environment, respectively. The dimensions of the system and environment are $d_{\text{S}}$ and $d_{\text{E}}$, while the probe is taken to be a qubit, \emph{i.e.,} $\dim(\mathcal{H}_{\text{P}}) = 2$. The computational basis states of the probe are denoted by $\ket{0_{\text{P}}}$ and $\ket{1_{\text{P}}}$.\par
Let the probe, system, and environment be initially prepared in the product state $\rho(0) = \ketbra{\chi_{\text{P}}}{\chi_{\text{P}}} \otimes \rho_{\text{S}} \otimes \ketbra{\xi_{\text{E}}}{\xi_{\text{E}}}$. We consider the joint unitary evolution operator for the probe-system-environment as
\begin{align}
\!\!\!U_{\text{\scriptsize{PSE}}}=\ketbra{0_{\text{P}}}{0_{\text{P}}}\otimes U_{\text{S}} \otimes\! I_{\text{E}} + \ketbra{1_{\text{P}}}{1_{\text{P}}} \otimes U_{\text{SE}} \!\circ\!(\widetilde{U}_{\text{S}} \otimes I_{\text{E}}), \label{FW-1}
\end{align}
where $U_{\text{S}}$ and $\widetilde{U}_{\text{S}}$ are distinct unitary operators acting on the system, and $U_{\text{SE}}$ is a unitary interaction between the system and the environment.
\begin{figure}[H]
\centering
\includegraphics[scale=0.46]{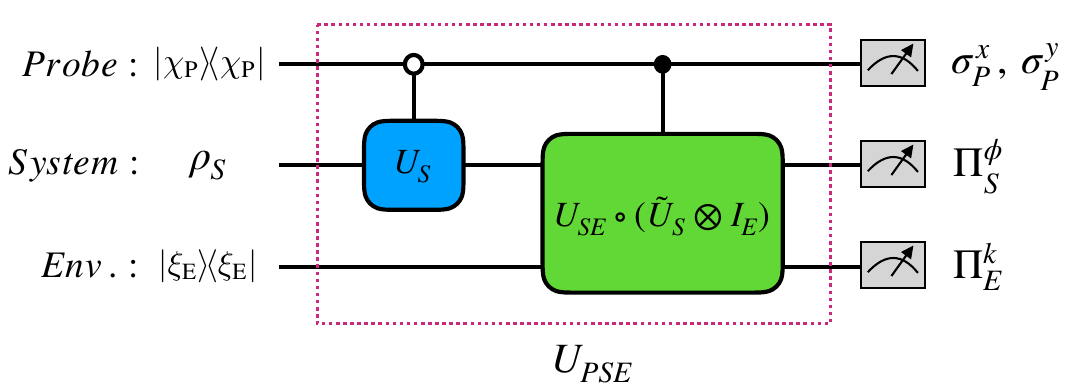}
\caption{Quantum circuit for implementation of the unitary operator $U_{\text{PSE}}$.}
\label{FIG1}
\end{figure}
Equation \eqref{FW-1} thus defines a controlled unitary evolution conditioned on the probe state:
\begin{itemize}[leftmargin=*, itemsep=-2pt, topsep=3pt]
\item  If the probe is in $\ket{0_{\text{P}}}$, only the system evolves under the unitary operator $U_{\text{S}}$.\\
\item  If the probe is in $\ket{1_{\text{P}}}$, the system-environment evolves under the unitary operator $U_{\text{SE}}\circ(\widetilde{U}_{\text{S}}\otimes I_{\text{E}})$. This is illustrated in Fig. \ref{FIG1}.
\end{itemize} 
After applying the unitary operator $U_{\text{PSE}}$, the initial product state $\rho(0)$ evolves as
\begin{align}
\rho(t) = U_{\text{PSE}}\, \rho(0)\, U_{\text{PSE}}^{\dagger}. \label{FW-2}
\end{align}
We then perform measurements of the Pauli operators $\sigma^x_{\text{P}}$ and $\sigma^y_{\text{P}}$ on the probe, projective measurements $\Pi^{\phi}_{\text{S}} = \ketbra{\phi_{\text{S}}}{\phi_{\text{S}}}$ on the system, and $\Pi^k_{\text{E}} = \ketbra{k_{\text{E}}}{k_{\text{E}}}$ on the environment. This yields the central result of this work:
\begin{align}
\braket{\phi_{\text{S}} | A_k \widetilde{U}_{\text{S}} \rho_{\text{S}} U_{\text{S}}^{\dagger} | \phi_{\text{S}}} 
= \frac{ \braket{ (\sigma^x_{\text{P}} + i \sigma^y_{\text{P}}) \otimes \Pi^{\phi}_{\text{S}} \otimes \Pi^{k}_{\text{E}} }_{\rho(t)} }{ \mathcal{N}_{\text{PSE}} }, \label{FW-3}
\end{align}
where $A_k = \braket{k_{\text{E}} | U_{\text{SE}} | \xi_{\text{E}}}$ is a Kraus operator, $\mathcal{N}_{\text{PSE}} = 2 \braket{\chi_{\text{P}} | 0_{\text{P}}} \braket{1_{\text{P}} | \chi_{\text{P}}} \braket{\xi_{\text{E}} | k_{\text{E}}}$, and $\braket{ (\sigma^x_{\text{P}} + i \sigma^y_{\text{P}}) \otimes \Pi^{\phi}_{\text{S}} \otimes \Pi^{k}_{\text{E}} }_{\rho(t)}$ is the average value of the tripartite operator $(\sigma^x_{\text{P}} + i \sigma^y_{\text{P}}) \otimes \Pi^{\phi}_{\text{S}}\otimes \Pi^{k}_{\text{E}}$ \emph{w.r.t} the time evolved tripartite density operator $\rho(t)$ given by Eq. \eqref{FW-2}. A detailed derivation of Eq.~(\ref{FW-3}) is provided in the Supplemental Material \cite{Supplemental}. This result is exact, \emph{i.e.}, it involves no approximations. In what follows, we demonstrate how Eq.~(\ref{FW-3}) can be used to extract matrix elements of a Kraus operator, a density matrix, a unitary operator, and an observable. Notably, it also enables the determination of the well-known weak value and modular value—without invoking the weak coupling approximation.
\vspace{3mm}
\\
\emph{Characterization of a Kraus operator.}--- To determine the $(i,j)$-th element of the Kraus operator $A_k$, we set $\widetilde{U}_\text{S} = I_\text{S}$, $\ket{\phi_\text{S}} = \ket{i}$, and $\rho_\text{S}=\ketbra{\psi_\text{S}}{\psi_\text{S}}=\ketbra{j}{j}$ in Eq.~(\ref{FW-3}), yielding
\begin{align}
\braket{i | A_k | j} = \frac{ \braket{ (\sigma^x_\text{P} + i \sigma^y_\text{P}) \otimes \Pi^i_\text{S} \otimes \Pi^k_\text{E} }_{\rho(t)} }{ \mathcal{N}_\text{PSE} \braket{j | U_\text{S}^\dagger | i} }. \label{DFCKO-1}
\end{align}
Here, $U_\text{S}$ is assumed to be a known unitary satisfying $\braket{j | U_\text{S}^\dagger | i} \neq 0$ for the specified $i,j$. The states $\ket{\chi_\text{P}}$ and $\ket{\xi_\text{E}}$ are chosen such that $\braket{\chi_\text{P} | 0_\text{P}} \neq 0$, $\braket{1_\text{P} | \chi_\text{P}} \neq 0$, and $\braket{\xi_\text{E} | k_\text{E}} \neq 0$, ensuring $\mathcal{N}_\text{PSE} \neq 0$.\par
To fully reconstruct $A_k$, let the input states be $\{\ket{j}\}_{j=0}^{d_\text{S}-1}$, forming a complete orthonormal basis of the system Hilbert space, such that $\sum_j \ketbra{j}{j} = I$. For each input $\ket{j}$, we apply the unitary $U_\text{PSE}$ [Eq.~(\ref{FW-1})] with $\widetilde{U}_\text{S} = I_\text{S}$ and measure an observable on the system whose eigenvectors are exactly  $\{\ket{\phi_{\text{S}}}\}=\{\ket{i}\}_{i=0}^{d_{\text{S}}-1}$ and on the probe using $\sigma^x_\text{P}$ and $\sigma^y_\text{P}$.\par 
A suitable choice for $U_\text{S}$ is the $d$-dimensional Hadamard gate, which satisfies $\braket{j | U_\text{S}^\dagger | i} \neq 0$ for all $i,j = 0,\ldots,d_\text{S}-1$~\cite{Sanders_Gate-2020}. Alternatively, one can construct $U_\text{S}$ using a projector $\Pi^{b^0}_\text{S} = \ketbra{b^0_\text{S}}{b^0_\text{S}}$, where $\ket{b^0_\text{S}} = \frac{1}{\sqrt{d_\text{S}}} \sum_{l=0}^{d_\text{S}-1} \ket{l}$ is a maximally coherent state, such that $\braket{j | e^{i \tilde{\theta} \Pi^{b^0}_\text{S}} | i} \neq 0$ for all $i,j$, where $\tilde{\theta}$ is a system parameter. With this procedure, each column of $A_k$ can be reconstructed, enabling full characterization of the Kraus operator $A_k$. A detailed implementation of our protocol using a modified Mach–Zehnder interferometer is provided in the Supplemental Material~\cite{Supplemental}.\par
Once the full matrix form of the Kraus operator $A_k$ is obtained, the corresponding POVM element follows directly as $E_k = A_k^{\dagger} A_k$. We now compare our method for characterizing $E_k$ with the recent approach in Ref.~\cite{Xu-2021-DirectMeasurements}. In that method, the determination of $(i,j)$-th element of $E_k$ requires $d_\text{S}$ distinct input states forming a complete basis $\{ \ket{n} \}_{n=0}^{d_\text{S} - 1}$, and one probe-system unitary of the form $e^{-i \theta \sigma^y_\text{P} \otimes \Pi^j_\text{S}}$, and measurement of the POVM element $E_k$. For the determination of $E_k$, it requires $d_\text{S}$ distinct input states forming a complete basis, and $d_\text{S}$ probe-system unitary of the form $\{ e^{-i \theta \sigma^y_\text{P} \otimes \Pi^j_\text{S}} \}_{j=0}^{d_\text{S}-1}$. Here $\theta$ represents the strength of the coupling. Additionally, their scheme involves three fixed probe measurements: $\sigma^x_\text{P}$, $\sigma^y_\text{P}$, and $\sigma^z_\text{P}$.\par
By contrast, our method requires only a single fixed unitary $U_{\text{PSE}}$ for each input state and two fixed measurement operators ($\sigma^x_\text{P}$ and $\sigma^y_\text{P}$) on the probe. As our scheme avoids any approximations, it offers inherently higher accuracy than methods relying on weak assumptions. Moreover, our protocol requires fewer quantum operations—both in terms of the number of distinct unitaries and measurements—thus offering greater precision. In the Supplemental Material~\cite{Supplemental}, we provide a detailed analysis of the error in estimating both individual POVM elements $E_k$ and the complete set $E = \{E_k\}_{k=0}^{d_{\text{E}} - 1}$, and compare our results with those reported in Ref.~\cite{Xu-2021-DirectMeasurements}.\par
Therefore, if a complete set of input states $\{ \ket{\psi_\text{S}} \}=\{\ket{j}\}_{j=0}^{d_\text{S}-1}$ is experimentally accessible, our procedure for determining the full Kraus operator $A_k$, and hence the POVM element $E_k$, is both efficient and experimentally feasible. In the Supplemental Material~\cite{Supplemental}, we present an alternative approach for accessing the $(i,j)$-th matrix element of $A_k$, based on the use of the Pauli $X$-gate and its higher-order generalizations.
\vspace{3mm}
\\
\emph{Characterization of a density matrix.}--- For the characterization of an unknown density matrix $\rho_\text{S}$ of the system, the system-environment interaction can be neglected in the derivation of Eq. \eqref{FW-3}. To extract the $(i,j)$-th element of $\rho_\text{S}$, we set $U_\text{SE} = I_\text{SE}$ and $\widetilde{U}_\text{S} = I_\text{S}$ in Eq.~\eqref{FW-1}, and choose $\ket{\phi_\text{S}} = \ket{i}$ in Eq.~\eqref{FW-3}. This yields the direct expression
\begin{align}
\braket{i | \rho_\text{S} | j} = \frac{1}{\mathcal{N}_\text{PS}} \braket{ (\sigma^x_\text{P} + i \sigma^y_\text{P}) \otimes \Pi^i_\text{S} }_{\rho(t)}, \label{DFCDM-1}
\end{align}
where we have chosen $U_\text{S}^\dagger \ket{i} = \ket{j}$, and $\mathcal{N}_\text{PS} = 2 \braket{\chi_\text{P} | 0_\text{P}} \braket{1_\text{P} | \chi_\text{P}}$. A detailed implementation of our protocol using a modified Mach–Zehnder interferometer is provided in the Supplemental Material~\cite{Supplemental}.\par
For full characterization of $\rho_\text{S}$, we take $U_\text{S}$ to be the $n$th-order generalized Pauli-$X$ gate, \emph{i.e.}, $U_\text{S} = X^n_\text{S}$, defined by its action: $X^n_\text{S} \ket{i} = \ket{(i + n) \bmod d_\text{S}}$ and ${X^n_\text{S}}^\dagger \ket{i} = \ket{(i - n) \bmod d_\text{S}}$~\cite{Babazadeh-2017,Gao-Zeilinger-2019,Wang_2022,Meng-2024}, where $n \in \mathbb{N}$. As an illustrative example, consider a three-dimensional system with measurement basis $\{ \ket{\phi_\text{S}} \} = \{ \ket{0}, \ket{1}, \ket{2} \}$. For $U_\text{S} = X^0_\text{S} = I_\text{S}$, we obtain the diagonal elements $\braket{0|\rho_\text{S}|0}$, $\braket{1|\rho_\text{S}|1}$, and $\braket{2|\rho_\text{S}|2}$. For $U_\text{S} = X^1_\text{S}$, we obtain the off-diagonal elements $\braket{0|\rho_\text{S}|2}$, $\braket{1|\rho_\text{S}|0}$, and $\braket{2|\rho_\text{S}|1}$. Thus, with only two unitaries—$X^0_\text{S}$ and $X^1_\text{S}$—we access six independent matrix elements, enabling efficient reconstruction of $\rho_\text{S}$ in this three-dimensional system. A detailed description of our method for general $d_\text{S}$-dimensional systems, along with examples, is provided in the Supplemental Material~\cite{Supplemental}.\par
We now compare our approach with the recent method proposed in Ref.~\cite{Vallone-2018}. That scheme requires $d_\text{S}$ distinct unitary operations acting on a tripartite system involving two probes and the system, along with three joint probe measurements. In contrast, our method requires only $(d_\text{S}/2 + 1)$ unitaries for even $d_\text{S}$ and $[(d_\text{S} - 1)/2 + 1]$ unitaries for odd $d_\text{S}$, all acting solely on the system; see Supplemental Material \cite{Supplemental}. Additionally, only two fixed measurements—$\sigma^x_\text{P}$ and $\sigma^y_\text{P}$—are needed on the probe. As our protocol involves no approximations, it offers inherently higher accuracy than methods based on weak assumptions. Moreover, our protocol requires fewer quantum operations—both in terms of the number of distinct unitaries and measurements—thus offering greater precision. In the Supplemental Material~\cite{Supplemental}, we present a detailed analysis of the error in estimating the unknown density matrix $\rho_{\text{S}}$, and compare our results with those of Refs.~\cite{Vallone-2018,Xu-Zhou-2024}. \par
Now we provide the characterization of a unitary operator which is essential for both weak value/modular value determination and observable characterization.
\vspace{3mm} 
\\
\emph{Characterization of a unitary operator.}--- Equation \eqref{FW-3} enables the characterization of an unknown unitary operator $\widetilde{U}_\text{S}$ also. We set $U_\text{SE} = I_\text{SE}$, and choose $\rho_\text{S}=\ketbra{\psi_\text{S}}{\psi_\text{S}}$ in Eq.~\eqref{FW-3}. This yields:
\begin{align}
\braket{\phi_{\text{S}}|\widetilde{U}_{\text{S}}|\psi_{\text{S}}} 
= \frac{ \braket{ (\sigma^x_{\text{P}} + i \sigma^y_{\text{P}}) \otimes \Pi^{\phi}_{\text{S}} }_{\rho(t)} }{ \mathcal{N}_{\text{PS}}\braket{\psi_{\text{S}}|U_{\text{S}}^{\dagger}|\phi_{\text{S}}} }, \label{DFCUO-1}
\end{align}
where $\mathcal{N}_{\text{PS}} = 2 \braket{\chi_{\text{P}} | 0_{\text{P}}} \braket{1_{\text{P}} | \chi_{\text{P}}}$. Let $\ket{\psi_{\text{S}}}=\ket{j}$, and $\ket{\phi_{\text{S}}}=\ket{i}$. Here, $U_\text{S}$ is assumed to be a known unitary satisfying $\braket{j | U_\text{S}^\dagger | i} \neq 0$ for the specified $i,j$. The states $\ket{\chi_\text{P}}$ and $\ket{\xi_\text{E}}$ are chosen such that $\braket{\chi_\text{P} | 0_\text{P}} \neq 0$, $\braket{1_\text{P} | \chi_\text{P}} \neq 0$, ensuring $\mathcal{N}_\text{PS} \neq 0$. Under the above conditions, we can obtain the $(i,j)$-th element of $\widetilde{U}_\text{S}$ \emph{i.e.,} $\braket{i|\widetilde{U}_{\text{S}}|j}$. The same procedure described for characterizing a Kraus operator can be followed here to fully characterize the unitary operator $\widetilde{U}_\text{S}$.
\vspace{3mm} 
\\
\emph{Determination of weak and modular values.---} The concept of the weak value of an observable was originally introduced by Aharonov, Albert, and Vaidman~\cite{Aharonov-Albert-Vaidman-1988} to investigate time asymmetry in quantum mechanics~\cite{Aharonov-Bergmann-Lebowitz-1964}. For detailed discussions on the foundational significance and diverse applications of weak values, including derivations in various contexts, see Refs.~\cite{Magana-Loaiza-2014, Hosten-Kwiat-2008, Aharonov-2013, Denkmayr-2014, Sahil-2023-URPPS, Aharonov-2002, Rozema-2012, Sahil-2023-WeakProduct}. A closely related quantity, known as the \emph{modular value}, was introduced in Ref.~\cite{Kedem-Vaidman-2010}, and shares a structural similarity with the weak value. Modular values have since been employed in a variety of contexts, including the derivation of nonlocal weak values and weak probabilities relevant to foundational phenomena such as the EPR paradox, Hardy’s paradox, and the quantum Cheshire Cat effect~\cite{Resch-Lundeen-Steinberg-2024, Ho-Imoto-2016-ModularFinite}. In what follows, we determine both the weak and modular values of observables and Kraus operators directly from Eq.~\eqref{DFCUO-1}, without invoking the weak measurement approximation.\par
\emph{Weak value}:---  In their framework \cite{Aharonov-Albert-Vaidman-1988}, weak values are accessed via weak interactions between the system and a probe. Here we show that our approach does not rely on any weak interaction approximations.\par
We consider the unitary operator $\widetilde{U}_{\text{S}}$ in Eq.~(\ref{DFCUO-1}) to have the form $\widetilde{U}_{\text{S}} = \widetilde{U}_{\text{S}}^{1} \circ \widetilde{U}_{\text{S}}^{2}$, where $\widetilde{U}_{\text{S}}^1 = e^{-i\theta_1 A}$ and $\widetilde{U}_{\text{S}}^2 = e^{i\theta_2 A}$, and $A$ is an observable whose weak value is to be determined. Then we have the following 
\begin{align}
\!\!\braket{\phi_{\text{S}}|\widetilde{U}_{\text{S}}|\psi_{\text{S}}} &= \braket{\phi_{\text{S}}|e^{-i(\theta_1 - \theta_2)A}|\psi_{\text{S}}} \nonumber\\
&\approx \braket{\phi_{\text{S}}|\psi_{\text{S}}} \left[ 1 - i(\theta_1 - \theta_2) \frac{\braket{\phi_{\text{S}}|A|\psi_{\text{S}}}}{\braket{\phi_{\text{S}}|\psi_{\text{S}}}} \right], \label{WV-1}
\end{align}
where we assumed $\theta_1 \approx \theta_2$ but explicitly not $\theta_{1,2} \approx 0$; that is, we do not assume weak interactions. Under $\theta_1 \approx \theta_2$, second and higher-order terms in the expansion vanish. From Eq.~(\ref{WV-1}), the weak value of $A$ can be extracted as
\begin{align}
\braket{A_w}^{\phi_{\text{S}}}_{\psi_{\text{S}}} := \frac{\braket{\phi_{\text{S}}|A|\psi_{\text{S}}}}{\braket{\phi_{\text{S}}|\psi_{\text{S}}}} = \frac{1}{i\delta\theta} \left[ 1 - \frac{\braket{\phi_{\text{S}}|\widetilde{U}_{\text{S}}|\psi_{\text{S}}}}{\braket{\phi_{\text{S}}|\psi_{\text{S}}}} \right], \label{WV-2}
\end{align}
where $\delta\theta = \theta_1 - \theta_2$, and it is assumed that \(\braket{\phi_{\text{S}}|\psi_{\text{S}}} \neq 0\). By substituting the expression for $\braket{\phi_{\text{S}}|\widetilde{U}_{\text{S}}|\psi_{\text{S}}}$ with $U_{\text{S}} = I_{\text{S}}$ from Eq.~(\ref{DFCUO-1}) into Eq.~(\ref{WV-2}), one can compute the weak value of $A$. Notably, this approach does not rely on a weak interaction regime for $\theta_1$ and $\theta_2$. By selecting $\theta_1$ and $\theta_2$ arbitrarily close, a near-ideal weak value can be obtained. In the Supplemental Material \cite{Supplemental}, we show that the accuracy of our weak value determination can be improved either by discarding third- and higher-order terms, or by employing an exact method that avoids any approximations using the notion of \textit{Modular Value}, which we discuss subsequently. Although Ref. \cite{Ogawa_2019} demonstrated that weak values can be obtained without invoking the weak approximation, their approach relies on a non-unitary probe-system transformation with restrictions on the observable and the interaction coefficient, which complicates experimental implementation. In contrast, our method employs a unitary transformation, offering a more practical alternative. We also note that our approach partially resembles the recent scheme proposed in Ref. \cite{Yoo_2025}.\par
It is also worth noting that the weak value of a Kraus operator \( A_k \) can be readily obtained by setting \( U_{\text{S}} = I \) in Eq.~(\ref{DFCKO-1}), yielding \( \braket{(A_k)_w}^{\phi_{\text{S}}}_{\psi_{\text{S}}} := \frac{\braket{\phi_{\text{S}}|A_k|\psi_{\text{S}}}}{\braket{\phi_{\text{S}}|\psi_{\text{S}}}} \). These weak values may be viewed as a generalization of those associated with orthogonal projection operators and can serve similar interpretational and practical roles. For instance, our results may offer new insights into the existence of so-called ``\textit{quantum Cheshire cats}," shedding light on how a quantum particle may be considered distinct from its properties—such as spin or polarization—in open quantum systems~\cite{Aharonov-2013,Denkmayr-2014}. Furthermore, the concept of the ``\emph{past of a particle}''---where one infers a particle’s presence through its nontrivial weak traces---can be naturally extended using nonzero weak values of Kraus operators \cite{Vaidman-Past-2013,Englert-Past-2017,Hance-2023}. This approach is more general than previous analyses based solely on projectors, as Kraus operators encompass both projective and non-projective measurements.\par
\emph{Modular value}:---
One can directly compute the modular value of the observable \( A \) using Eq.~(\ref{DFCUO-1}) by substituting \(\widetilde{U}_{\text{S}} = e^{-i\theta A}\) and \(U_{\text{S}} = I_{\text{S}}\), yielding
\begin{align}
\braket{A_m}_{\psi_{\text{S}}}^{\phi_{\text{S}}} := \frac{\braket{\phi_{\text{S}}|e^{-i\theta A}|\psi_{\text{S}}}}{\braket{\phi_{\text{S}}|\psi_{\text{S}}}}, \label{MV-1}
\end{align}
where \(|\psi_{\text{S}}\rangle\) and \(|\phi_{\text{S}}\rangle\) denote the pre- and postselected states, respectively.
It is assumed that \(\braket{\phi_{\text{S}}|\psi_{\text{S}}} \neq 0\).\par
In Ref.~\cite{Kedem-Vaidman-2010}, the authors derived the modular value by considering a probe-system Hamiltonian of the form 
\begin{align*}
H_{\text{PS}}^{\text{Kedem-Vaidman}} = g\ketbra{1_{\text{P}}}{1_{\text{P}}} \otimes A.
\end{align*}
In contrast, our formulation in Eq.~(\ref{MV-1}) incorporates contributions from both pointer states \(\ket{0_{\text{P}}}\) and \(\ket{1_{\text{P}}}\), indicating that the modular value in our scheme is obtained without disregarding any pointer outcome. Moreover, unlike the approach of Ref.~\cite{Kedem-Vaidman-2010}, our method does not require full state tomography of the probe at the end of the process. Specifically, as can be seen from Eq.~(\ref{MV-1}), our scheme requires only measurements of the Pauli operators \(\sigma_\text{S}^x\) and \(\sigma^y_\text{S}\) on the probe. In comparison, the method in Ref.~\cite{Kedem-Vaidman-2010} needs measurements of \(\sigma^x_\text{S}\), \(\sigma^y_\text{S}\), and \(\sigma^z_\text{S}\) to reconstruct the full qubit probe state.
\vspace{2mm}
\\
\emph{Direct characterization of an unknown observable.---}
We now discuss two different approaches for obtaining the matrix elements of an unknown observable \( A \), depending on the experimental contexts.\par
\emph{First scenario}: Suppose the experimental setup allows one to engineer the probe-system interaction using the projection operators of the unknown observable \( A \). In this case, we have $\widetilde{U}_{\text{S}}=e^{-i \theta\Pi^{a_k}_\text{S}}$, where \(\{ \ket{a_k} \}\) are the eigenstates of \( A \). Then one obtains the matrix elements of $\widetilde{U}_{\text{S}}$ as
\begin{align*}
\braket{i|\widetilde{U}_{\text{S}}|j} = \braket{i|e^{-i \theta\Pi^{a_k}_{\text{S}}}|j} = \delta_{ij} + (e^{-i t} - 1)\braket{i|\Pi_{\text{S}}^{a_k}|j}.
\end{align*}
This directly yields the matrix element \(\braket{i|\Pi_{\text{S}}^{a_k}|j}\) by substituting the matrix element $\braket{i|\widetilde{U}_{\text{S}}|j}$  given in Eq.~(\ref{DFCUO-1}). Once all such elements are obtained for the set \(\{ \Pi_{\text{S}}^{a_k} \}\), the matrix elements of \( A \) follow from the spectral decomposition:
\begin{align*}
\braket{i|A|j} = \sum_k a_k \braket{i|\Pi_{\text{S}}^{a_k}|j},
\end{align*}
provided that the eigenvalues \(\{a_k\}\) of \( A \) are known. \par
\emph{Second scenario}: If the experimental setup does not allow for the preparation of the probe-system interaction using the projection operators of the unknown observable A—a situation that may arise in weak interaction regimes—an alternative method can be employed. In this case, the matrix element \(\braket{i|A|j}\) can be expanded using the identity, \( I_{\text{S}} = \sum_{k=0}^{d_{\text{S}}-1} \ket{\lambda_k}\!\bra{\lambda_k} \), as
\begin{align*}
\braket{i|A|j} = \sum_{k=0}^{d_{\text{S}}-1} \braket{i|\lambda_k} \braket{\lambda_k|j} \braket{A_w}_{j}^{\lambda_k},
\end{align*}
where the weak value \(\braket{A_w}_{j}^{\lambda_k}\) is defined in Eq.~(\ref{WV-2}). This decomposition holds under the conditions \(\braket{i|\lambda_k} \neq 0\) and \(\braket{\lambda_k|j} \neq 0\). Each weak value \(\braket{A_w}_{j}^{\lambda_k}\) can be experimentally determined using the procedure outlined in Eq.~(\ref{WV-2}).
\vspace{2mm}
\\
\emph{Discussion and Conclusion}.--- 
In prior approaches to the direct characterization of quantum measurements, the primary focus has been on POVMs, rather than the Kraus operators that generate them. Characterizing a POVM element alone is insufficient to fully capture the underlying physical dynamics, as different sets of Kraus operators can produce identical measurement statistics. This creates a crucial gap between knowledge of the measurement outcomes and the physical mechanisms responsible for them. In this work, we have addressed this gap by presenting a framework for the direct characterization of individual Kraus operators. Our method also enables the direct reconstruction of density matrices by determining their individual elements. Our conceptually simple protocol has also been used for determining modular and weak values. A notable extension is its applicability to projective measurements and observables, enabling the reconstruction of unknown projective measurements and observables through their matrix elements via weak values. Furthermore, we have shown that the framework naturally extends to the characterization of unitary operators. We have also discussed the foundational and interpretational implications of weak values of Kraus operators—viewed as a generalization of the weak values of projection operators—in contexts such as the ``\textit{Quantum Cheshire Cat}" phenomenon and the ``\emph{past of a particle}.'' The out-of-time-ordered correlators (OTOCs) can also be computed within our framework for arbitrary mixed states, as demonstrated in \cite{Swingle-OTOC-2016}.\par
The key advantage of our approach is its independence from the weak coupling approximation. Unlike existing DCQM and DCDM protocols, which typically rely on weak (or strong but restricted) probe-system couplings, or require detailed modeling of complex probe-system-environment interactions and the $d_\text{S}$ number of unitary evolution operators in terms of projectors, our method remains valid across arbitrary interaction strengths. It does not require detailed modeling of the probe-system-environment dynamics and involves fewer quantum operations—both in terms of the number of distinct unitaries and measurements. These features enhance both the precision and the practical applicability of our framework.\par
Overall, we have introduced a unified framework that integrates and extends previous techniques based on weak values, modular values, and direct characterization schemes, while remaining experimentally feasible across a variety of platforms—including optical interferometry \cite{Babazadeh-2017,Gao-Zeilinger-2019,Wang_2022,Meng-2024}, superconducting qubits \cite{Yamamoto-2003,Galiautdinov-2007,Kandala-2021}, trapped-ion systems \cite{Monz-2009,Tan-2015}, and so on. As demonstrated in the Supplemental Material \cite{Supplemental}, our protocol can be implemented using modified Mach--Zehnder interferometers. \par
In future, our framework will provide a promising foundation for quantum information tasks that require precise characterization of measurements, such as efficient tracking of open quantum system dynamics under continuous monitoring. Its reliance on structurally simple evolution unitaries and compatibility with current experimental technologies position it as a versatile tool for the direct characterization of Kraus operators, density matrices, observables, and unitary operations, as well as average and weak values in multipartite quantum systems.
\vspace{2mm}
\\
\emph{Acknowledgments.---} Sahil gratefully acknowledges Prof. Sibasish Ghosh for insightful discussions.

\bibliography{main}

\clearpage

\onecolumngrid
\section{SUPPLEMENTAL MATERIAL}

\section{S1. Derivation of Eq. \eqref{FW-3}}
Let the probe, system, and environment be initially prepared in the product state $\rho(0) = \ketbra{\chi_{\text{P}}}{\chi_{\text{P}}} \otimes \rho_{\text{S}} \otimes \ketbra{\xi_{\text{E}}}{\xi_{\text{E}}}$. The joint unitary evolution operator for the probe-system-environment is given in Eq. \eqref{FW-1} as
\begin{align*}
\!\!\!U_{\text{\scriptsize{PSE}}}=\ketbra{0_{\text{P}}}{0_{\text{P}}}\otimes U_{\text{S}} \otimes\! I_{\text{E}} + \ketbra{1_{\text{P}}}{1_{\text{P}}} \otimes U_{\text{SE}} \!\circ\!(\widetilde{U}_{\text{S}} \otimes I_{\text{E}}).
\end{align*}
After applying the unitary operator $U_{\text{PSE}}$, the initial product state $\rho(0)$ evolves as
\begin{align*}
\rho(t) = U_{\text{PSE}}\, \rho(0)\, U_{\text{PSE}}^{\dagger}.
\end{align*}
We then perform measurements of the Pauli operators $\sigma^x_{\text{P}}=\ketbra{+_{\text{P}}}{+_{\text{P}}}-\ketbra{-_{\text{P}}}{-_{\text{P}}}$ and $\sigma^y_{\text{P}}=\ketbra{+i_{\text{P}}}{+i_{\text{P}}}-\ketbra{-i_{\text{P}}}{-i_{\text{P}}}$ on the probe, projective measurements $\Pi^{\phi}_{\text{S}} = \ketbra{\phi_{\text{S}}}{\phi_{\text{S}}}$ on the system, and $\Pi^k_{\text{E}} = \ketbra{k_{\text{E}}}{k_{\text{E}}}$ on the environment. Now the probability that the probe-system-environment will be in the state $\ket{\widetilde{\chi}_{\text{P}}}\otimes\ket{\phi_{\text{S}}}\otimes\ket{k_{\text{E}}}$ is given by 
\begin{align}
p(\widetilde{\chi}_{\text{P}},\phi_{\text{S}},k_{\text{E}})&=Tr\left[\left(\Pi^{\widetilde{\chi}}_{\text{P}} \otimes \Pi^{\phi}_{\text{S}} \otimes \Pi^{k}_{\text{E}}\right)\rho(t)\right]\nonumber\\
&=|\braket{0_{\text{P}}|\chi_{\text{P}}}|^2|\braket{\widetilde{\chi}_{\text{P}}|0_{\text{P}}}|^2|\braket{\xi_{\text{E}}|k_{\text{E}}}|^2\braket{\phi_{\text{S}}|U_{\text{S}}\rho_{\text{S}}U_{\text{S}}^{\dagger}|\phi_{\text{S}}}\nonumber\\
& \quad+ \braket{0_{\text{P}}|\chi_{\text{P}}}\braket{\chi_{\text{P}}|1_{\text{P}}}\braket{\widetilde{\chi}_{\text{P}}|0_{\text{P}}}\braket{1_{\text{P}}|\widetilde{\chi}_{\text{P}}}\bra{\phi_{\text{S}}}\otimes\bra{k_{\text{E}}}\left[\left(U_{\text{S}}\rho_{\text{S}}\widetilde{U}_{\text{S}}^{\dagger}\otimes\ketbra{\xi_{\text{E}}}{\xi_{\text{E}}}\right)U_{\text{SE}}^{\dagger}\right]\ket{\phi_{\text{S}}}\otimes\ket{k_{\text{E}}}\nonumber\\
& \quad+ \braket{\chi_{\text{P}}|0_{\text{P}}}\braket{1_{\text{P}}|\chi_{\text{P}}}\braket{0_{\text{P}}|\widetilde{\chi}_{\text{P}}}\braket{\widetilde{\chi}_{\text{P}}|1_{\text{P}}}\bra{\phi_{\text{S}}}\otimes\bra{k_{\text{E}}}\left[U_{\text{SE}}\left(\widetilde{U}_{\text{S}}\rho_{\text{S}}U_{\text{S}}^{\dagger}\otimes\ketbra{\xi_{\text{E}}}{\xi_{\text{E}}}\right)\right]\ket{\phi_{\text{S}}}\otimes\ket{k_{\text{E}}}\nonumber\\
& \quad+ |\braket{1_{\text{P}}|\chi_{\text{P}}}|^2|\braket{\widetilde{\chi}_{\text{P}}|1_{\text{P}}}|^2 \bra{\phi_{\text{S}}}\otimes\bra{k_{\text{E}}}\left[U_{\text{SE}}\left(\widetilde{U}_{\text{S}}\rho_{\text{S}}\widetilde{U}_{\text{S}}^{\dagger}\otimes\ketbra{\xi_{\text{E}}}{\xi_{\text{E}}}\right)U_{\text{SE}}^{\dagger}\right]\ket{\phi_{\text{S}}}\otimes\ket{k_{\text{E}}},\tag{S1-1}\label{S1-1}
\end{align}
where $\ket{\widetilde{\chi}_{\text{P}}}\in\{\ket{+_{\text{P}}},\ket{-_{\text{P}}},\ket{+i_{\text{P}}},\ket{-i_{\text{P}}}\}$. Now consider the following:
\begin{align}
\bra{\phi_{\text{S}}}\otimes\bra{k_{\text{E}}}\left[U_{\text{SE}}\left(\widetilde{U}_{\text{S}}\rho_{\text{S}}U_{\text{S}}^{\dagger}\otimes\ketbra{\xi_{\text{E}}}{\xi_{\text{E}}}\right)\right]\ket{\phi_{\text{S}}}\otimes\ket{k_{\text{E}}}&=\braket{\xi_{\text{E}}|k_{\text{E}}}\bra{\phi_{\text{S}}}\left[\braket{k_{\text{E}}|U_{\text{SE}}|\xi_{\text{E}}}\widetilde{U}_{\text{S}}\rho_{\text{S}}U_{\text{S}}^{\dagger}\right]\ket{\phi_{\text{S}}}\nonumber\\
&=\braket{\xi_{\text{E}}|k_{\text{E}}}\braket{\phi_{\text{S}}|A_k\widetilde{U}_{\text{S}}\rho_{\text{S}}U_{\text{S}}^{\dagger}|\phi_{\text{S}}},\tag{S1-2}\label{S1-2}
\end{align}
where $A_k$ is the Kraus operator defined as $A_k := \braket{k_{\text{E}}|U_{\text{SE}}|\xi_{\text{E}}}$. By substituting Eq.~\eqref{S1-2} into Eq.~\eqref{S1-1}, and setting $\ket{\pm_{\text{P}}} = \frac{1}{\sqrt{2}}(\ket{0_{\text{P}}} \pm \ket{1_{\text{P}}})$ and $\ket{\pm i_{\text{P}}} = \frac{1}{\sqrt{2}}(\ket{0_{\text{P}}} \pm i\ket{1_{\text{P}}})$ in Eq.~\eqref{S1-2}, we obtain, after some manipulation,
\begin{align}
p(+_{\text{P}},\phi_{\text{S}},k_{\text{E}})-p(-_{\text{P}},\phi_{\text{S}},k_{\text{E}})=\braket{\sigma^x_{\text{P}} \otimes \Pi^{\phi}_{\text{S}} \otimes \Pi^{k}_{\text{E}}}_{\rho(t)}
&=2\text{Re}\left[\braket{\chi_{\text{P}}|0_{\text{P}}}\braket{1_{\text{P}}|\chi_{\text{P}}}\braket{\xi_{\text{E}}|k_{\text{E}}}\braket{\phi_{\text{S}}|A_k\widetilde{U}_{\text{S}}\rho_{\text{S}}U_{\text{S}}^{\dagger}|\phi_{\text{S}}}\right],\tag{S1-3}\label{S1-3}
\end{align}
\begin{align}
p(+i_{\text{P}},\phi_{\text{S}},k_{\text{E}})-p(-i_{\text{P}},\phi_{\text{S}},k_{\text{E}})=\braket{\sigma^y_{\text{P}} \otimes \Pi^{\phi}_{\text{S}} \otimes \Pi^{k}_{\text{E}}}_{\rho(t)}
&=2\text{Im}\left[\braket{\chi_{\text{P}}|0_{\text{P}}}\braket{1_{\text{P}}|\chi_{\text{P}}}\braket{\xi_{\text{E}}|k_{\text{E}}}\braket{\phi_{\text{S}}|A_k\widetilde{U}_{\text{S}}\rho_{\text{S}}U_{\text{S}}^{\dagger}|\phi_{\text{S}}}\right].\tag{S1-4}\label{S1-4}
\end{align}
By combining Eqs. \eqref{S1-3} and \eqref{S1-4}, we obtain our main result given in Eq. (\ref{FW-3}).
\vspace{6cm}

\section{S2. Direct characterization of an unknown \emph{Kraus operator} in a modified Mach–Zehnder interferometer (MZI) using our method given in Eq. \eqref{DFCKO-1}}
\begin{figure}[H]
\centering
\includegraphics[scale=0.53]{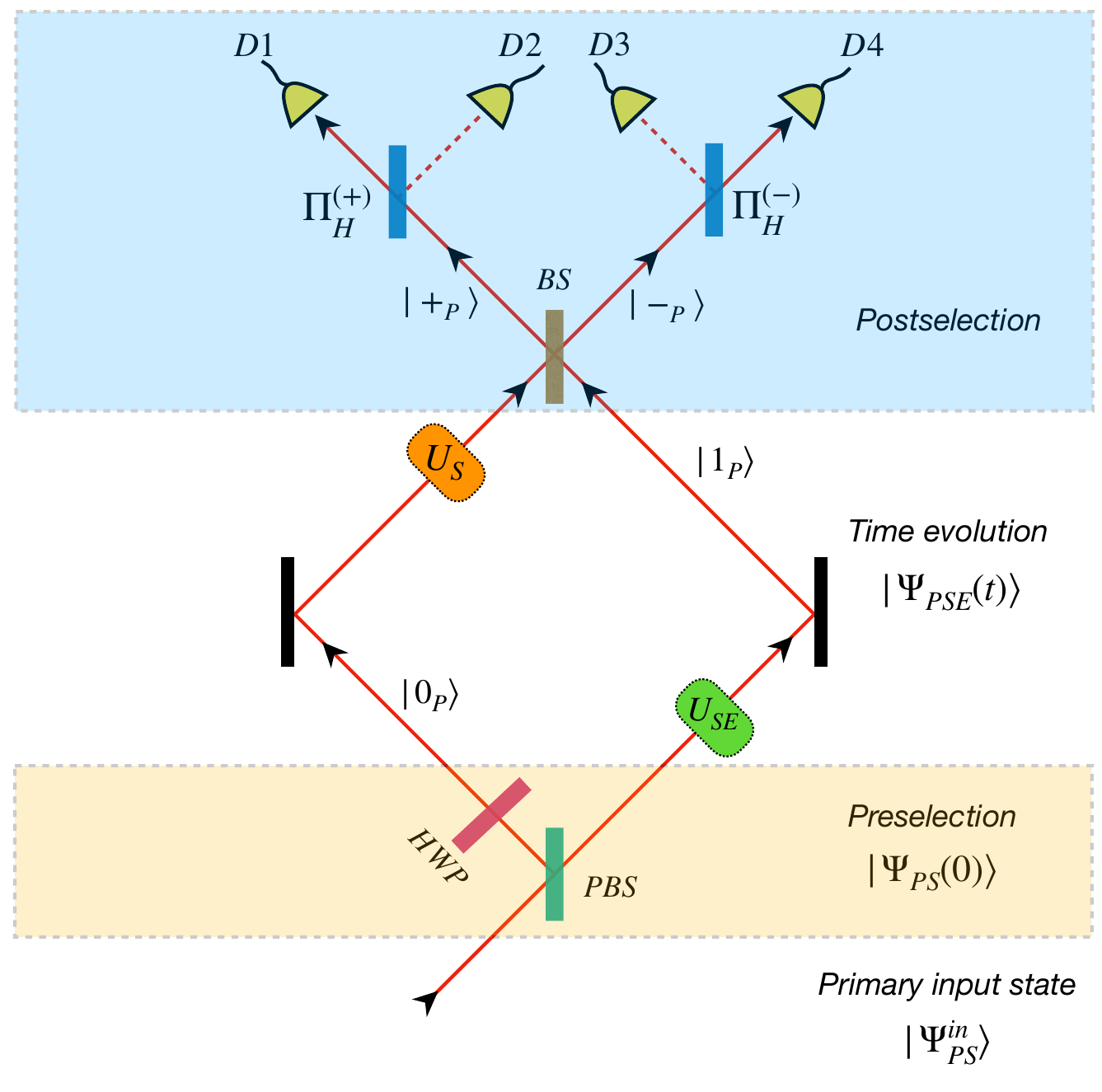}
\caption{Schematic illustration of a modified Mach–Zehnder interferometer (MZI) used for the direct characterization of an unknown Kraus operator.}
\label{FIG2}
\end{figure}
To realize our method of direct characterization of an unknown Kraus operator, we consider an experimental setup as depicted in Fig. \ref{FIG2}, which is a modified version of the standard Mach–Zehnder interferometer adapted for quantum optical implementation. Such setups have been extensively used in Refs. \cite{Aharonov-2013,Denkmayr-2014,Englert1996,Rab-Polino-Man-2017,Chowdhury-Pati-Chen-2021}. In this arrangement, a single photon is used, where the photon's path degree of freedom serves as the probe, and its polarization degree of freedom represents the system. The environment is considered as spin-$1/2$ particles, \emph{i.e.,} qubits.\par
The basis states for the probe, system, and environment are denoted as $\{\ket{0_{\text{P}}}, \ket{1_{\text{P}}}\}$, $\{\ket{H}, \ket{V}\}$, and $\{\ket{0_{\text{E}}}, \ket{1_{\text{E}}}\}$, respectively. Our goal is to determine the matrix element $\braket{H|A_0|V}$, where the Kraus operator $A_0 = \braket{0_{\text{E}}|U_{\text{SE}}|\xi_{\text{E}}}$ is defined in Eq.~(\ref{FW-3}).\par
Following the theoretical procedure from the main text, we begin by preparing the joint initial state:
\begin{align*}
\ket{\Psi_{\text{PS}}(0)} \otimes\ket{\xi_{\text{E}}} = (\cos\alpha\ket{0_{\text{P}}} + \sin\alpha\ket{1_{\text{P}}}) \otimes \ket{V} \otimes \ket{\xi_{\text{E}}},
\end{align*}
where the environment is initially in a pure uncorrelated state $\ket{\xi_{\text{E}}} = \cos\delta\ket{0_{\text{E}}} + \sin\delta\ket{1_{\text{E}}}$.\par
To achieve this, we start with the photon's primary input state $\ket{\Psi^{\text{in}}_{\text{PS}}} = \ket{a_{\text{P}}} \otimes (\cos\alpha\ket{H} + \sin\alpha\ket{V})$. Upon passing through a polarizing beam splitter (PBS), the state transforms as:
\begin{align*}
\ket{\Psi^{\text{in}}_{\text{PS}}} \xrightarrow{PBS} \cos\alpha \ket{0_{\text{P}}} \otimes \ket{H} + \sin\alpha \ket{1_{\text{P}}} \otimes \ket{V},
\end{align*}
creating entanglement between path (probe) and polarization (system). Next, a half-wave plate (HWP) is placed in path $\ket{0_{\text{P}}}$ to flip polarization: $\ket{H} \xrightarrow{HWP} \ket{V}$, resulting in the desired preselected state of the probe-system,
\begin{align*}
\ket{\Psi_{\text{PS}}(0)} = (\cos\alpha\ket{0_{\text{P}}} + \sin\alpha\ket{1_{\text{P}}}) \otimes \ket{V}.
\end{align*}
Before entering the evolution region of the modified MZI, the total state is $\ket{\Psi_{\text{PS}}(0)}\otimes\ket{\xi_{\text{E}}}$. Upon entering --- if the photon follows path $\ket{0_{\text{P}}}$ --- it evolves as:
\begin{align*}
\ket{0_{\text{P}}}\otimes \ket{V}\otimes \ket{\xi_{\text{E}}} \xrightarrow{I_{\text{P}} \otimes U_{\text{S}} \otimes I_{\text{E}}} \ket{0_{\text{P}}}\otimes U_{\text{S}} \ket{V}\otimes \ket{\xi_{\text{E}}},
\end{align*}
if it takes path $\ket{1_{\text{P}}}$, it undergoes:
\begin{align*}
\ket{1_{\text{P}}}\otimes \ket{V}\otimes \ket{\xi_{\text{E}}} \xrightarrow{I_{\text{P}} \otimes U_{\text{SE}}} \ket{1_{\text{P}}} \otimes U_{\text{SE}} \ket{V} \otimes\ket{\xi_{\text{E}}}.
\end{align*}
Thus, after the time evolution, the total state becomes:
\begin{align*}
\ket{\Psi_{\text{PSE}}(t)} = \cos\alpha \ket{0_{\text{P}}}\otimes U_{\text{S}} \ket{V}\otimes \ket{\xi_{\text{E}}} + \sin\alpha \ket{1_{\text{P}}}\otimes U_{\text{SE}} \ket{V}\otimes \ket{\xi_{\text{E}}}.
\end{align*}
The beam splitter (BS) is defined by:
\begin{align*}
\ket{0_{\text{P}}} \xrightarrow{BS} \frac{1}{\sqrt{2}}(\ket{+_{\text{P}}} + \ket{-_{\text{P}}}), \quad
\ket{1_{\text{P}}} \xrightarrow{BS} \frac{1}{\sqrt{2}}(\ket{+_{\text{P}}} - \ket{-_{\text{P}}}),
\end{align*}
so that the state after the BS becomes:
\begin{align*}
\ket{\Psi_{\text{PSE}}^{(BS)}(t)} &= \frac{1}{\sqrt{2}} \ket{+_{\text{P}}}\otimes[\cos\alpha\, U_{\text{S}} \ket{V} \otimes\ket{\xi_{\text{E}}} + \sin\alpha\, U_{\text{SE}} \ket{V} \otimes\ket{\xi_{\text{E}}}]\nonumber\\
&\quad + \frac{1}{\sqrt{2}} \ket{-_{\text{P}}}\otimes[\cos\alpha\, U_{\text{S}} \ket{V}\otimes \ket{\xi_{\text{E}}} - \sin\alpha\, U_{\text{SE}} \ket{V}\otimes \ket{\xi_{\text{E}}}].
\end{align*}
We define the postselection operator $\Pi_H^{(+)} = \ketbra{H}{H}$, which transmits only the horizontal polarization. Hence, if detector $D1$ clicks, we know that the photon is in the state $\ket{+_{\text{P}}} \otimes \ket{H}$. Immediately after, we measure the environment in the state $\ket{0_{\text{E}}}$. The probability of this joint detection is:
\begin{align}
p(+_{\text{P}}, H, 0_{\text{E}}) &= \frac{1}{2} \cos^2\alpha \cos^2\delta\, |\braket{H|U_{\text{S}}|V}|^2 + \frac{1}{2} \sin^2\alpha\, |\braket{H|A_0|V}|^2 \notag \\
&\quad + \text{Re}\big[\braket{V|U_{\text{S}}^\dagger|H} \braket{H|A_0|V}\big] \cos\alpha \sin\alpha \cos\delta. \tag{S2-1}\label{S2-1}
\end{align}
Similarly, for $\ket{-_{\text{P}}} \otimes \ket{H}$ and $\ket{0_{\text{E}}}$, the probability is:
\begin{align}
p(-_{\text{P}}, H, 0_{\text{E}}) &= \frac{1}{2} \cos^2\alpha \cos^2\delta\, |\braket{H|U_{\text{S}}|V}|^2 + \frac{1}{2} \sin^2\alpha\, |\braket{H|A_0|V}|^2 \notag \\
&\quad - \text{Re}\big[\braket{V|U_{\text{S}}^\dagger|H} \braket{H|A_0|V}\big] \cos\alpha \sin\alpha \cos\delta. \tag{S2-2}\label{S2-2}
\end{align}
Next, we replace the BS with a modified one, denoted by $\widetilde{BS}$, whose action is:
\begin{align*}
\ket{0_{\text{P}}} \xrightarrow{\widetilde{BS}} \frac{1}{\sqrt{2}}(\ket{+i_{\text{P}}} + i \ket{-i_{\text{P}}}), \quad
\ket{1_{\text{P}}} \xrightarrow{\widetilde{BS}} \frac{1}{\sqrt{2}}(\ket{+i_{\text{P}}} - i \ket{-i_{\text{P}}}).
\end{align*}
Here, we simply rename the paths by replacing $\ket{\pm_{\text{P}}}$ with $\ket{\pm i_{\text{P}}}$. Under this transformation, the probabilities for the photon being found in the $\ket{\pm i_{\text{P}}} \otimes \ket{H}$ states with the environment in $\ket{0_{\text{E}}}$ become:
\begin{align}
p(+i_{\text{P}}, H, 0_{\text{E}}) &= \frac{1}{2} \cos^2\alpha \cos^2\delta\, |\braket{H|U_{\text{S}}|V}|^2 + \frac{1}{2} \sin^2\alpha\, |\braket{H|A_0|V}|^2 \notag \\
&\quad + \text{Im}\big[\braket{V|U_{\text{S}}^\dagger|H} \braket{H|A_0|V}\big] \cos\alpha \sin\alpha \cos\delta, \tag{S2-3}\label{S2-3}
\\
p(-i_{\text{P}}, H, 0_{\text{E}}) &= \frac{1}{2} \cos^2\alpha \cos^2\delta\, |\braket{H|U_{\text{S}}|V}|^2 + \frac{1}{2} \sin^2\alpha\, |\braket{H|A_0|V}|^2 \notag \\
&\quad - \text{Im}\big[\braket{V|U_{\text{S}}^\dagger|H} \braket{H|A_0|V}\big] \cos\alpha \sin\alpha \cos\delta. \tag{S2-4}\label{S2-4}
\end{align}

Combining Eqs. (\ref{S2-1})–(\ref{S2-4}), the desired matrix element is reconstructed as:
\begin{align*}
\braket{H|A_0|V} &= \Big\{p(+_{\text{P}}, H, 0_{\text{E}}) - p(-_{\text{P}}, H, 0_{\text{E}})+ i \big[p(+i_{\text{P}}, H, 0_{\text{E}}) - p(-i_{\text{P}}, H, 0_{\text{E}})\big] \Big\} \times \frac{1}{2 \cos\alpha \sin\alpha \cos\delta\, \braket{V|U_{\text{S}}^\dagger|H}}.
\end{align*}

\section{S3. Precision in a POVM element Characterization: Our Method vs. Existing Approaches}
We assume that each matrix element $\braket{i|E_k|j}$  of a POVM element $E_k$ is a function of probabilities. \par
Now we define the error in estimating the matrix element $\braket{i|E_k|j}$ as
\begin{align}
|\delta\braket{i|E_k|j}|^2=\delta \text{Re}[\braket{i|E_k|j}]^2+\delta \text{Im}[\braket{i|E_k|j}]^2,\tag{S3-1}\label{S3-1}
\end{align}
where 
\begin{equation}\tag{S3-2}\label{S3-2}
\begin{aligned}
\delta \text{Re}[\braket{i|E_k|j}]^2&=\sum_{l\in \mathcal{M}} \left(\frac{\partial \text{Re}[\braket{i|E_k|j}]}{\partial p(l_{\text{P}},i_{\text{S}},k_{\text{E}})}\right)^2\delta^2p(l_{\text{P}},i_{\text{S}},k_{\text{E}}),\\
\delta \text{Im}[\braket{i|E_k|j}]^2&=\sum_{l\in \mathcal{M}} \left(\frac{\partial \text{Im}[\braket{i|E_k|j}]}{\partial p(l_{\text{P}},i_{\text{S}},k_{\text{E}})}\right)^2\delta^2p(l_{\text{P}},i_{\text{S}},k_{\text{E}}),
\end{aligned}
\end{equation}
$\delta^2p(l_{\text{P}},i_{\text{S}},k_{\text{E}})$ is the variance of the measured probability $p(l_{\text{P}},i_{\text{S}},k_{\text{E}})$, and  $\mathcal{M}$ is the set of quantum pure states. The error in estimating the unknown POVM element $E_k$ is defined as the sum of all the errors in estimating each matrix element $\braket{i|E_k|j}$ of $E_k$:
\begin{align}
\delta E_k=\sqrt{\sum_{i,j=0}^{d_{\text{S}}-1}|\delta\braket{i|E_k|j}|^2}.\tag{S3-3}\label{S3-3}
\end{align}
The error in estimating the POVM, $E=\{E_k\}_{k=0}^{d_{\text{E}}-1}$ is:
\begin{align}
\delta E=\sqrt{\sum_{k=0}^{d_{\text{E}}-1}\delta E_k^2}.\tag{S3-4}\label{S3-4}
\end{align}

\subsection{S3.1. By our method given in Eq. \eqref{DFCKO-1}}
In our method, as given in Eq.~\eqref{DFCKO-1}, we constructed the Kraus operator $A_k$ rather than the POVM element $E_k$. The latter is obtained using the relation $E_k = A_k^\dagger A_k$. Since the error in estimating $E_k$ is same as the error in estimating $A_k$, we begin by analyzing the error in $A_k$ first.\par
The real and imaginary parts of the element $\braket{i|A_k|j}$ are:
\begin{equation}\tag{S3.1-1}\label{S3.1-1}
\begin{aligned}
\text{Re}[\braket{i|A_k|j}]&=\frac{1}{\sqrt{\alpha_k}\braket{j|U_{\text{S}}^{\dagger}|i}}\left[p(+_{\text{P}},i_{\text{S}},k_{\text{E}})-p(-_{\text{P}},i_{\text{S}},k_{\text{E}})\right],\\
\text{Im}[\braket{i|A_k|j}]&=\frac{1}{\sqrt{\alpha_k}\braket{j|U_{\text{S}}^{\dagger}|i}}\left[p(+i_{\text{P}},i_{\text{S}},k_{\text{E}})-p(-i_{\text{P}},i_{\text{S}},k_{\text{E}})\right],
\end{aligned}
\end{equation}
where $p(\pm_{\text{P}},i_{\text{S}},k_{\text{E}})=Tr[(\ketbra{\pm_{\text{P}}}{\pm_{\text{P}}}\otimes\Pi^i_{\text{S}}\otimes\Pi_{\text{E}}^k)\rho(t)]$, $p(\pm i_{\text{P}},i_{\text{S}},k_{\text{E}})=Tr[(\ketbra{\pm i_{\text{P}}}{\pm i_{\text{P}}}\otimes\Pi^i_{\text{S}}\otimes\Pi_{\text{E}}^k)\rho(t)]$, and we have taken $\ket{\chi_{\text{P}}}=\frac{1}{\sqrt{2}}(\ket{0_{\text{P}}}+\ket{1_{\text{P}}}), \braket{\xi_{\text{E}}|k_{\text{E}}}=\sqrt{\alpha_k}\implies\mathcal{N}_{\text{PSE}}=\sqrt{\alpha_k}$ in Eq. \eqref{DFCKO-1}; here $\alpha_k>0$ and $\sum_k\alpha_k=1$. Also we assume that $U_{\text{S}}$ is a Hadamard matrix $H_{\text{S}}$ with $\braket{i|H^{\dagger}_{\text{S}}|j}$ being $\pm\frac{1}{\sqrt{d_{\text{S}}}}$ $\,\,$ $\forall\, i,j$. Note that $\{\ket{\pm_{\text{P}}}\}$ and $\{\ket{\pm i_{\text{P}}}\}$ are eigenstates of the Pauli operators $\sigma_{\text{P}}^x$ and $\sigma_{\text{P}}^y$, respectively.\par
Now we assume that $N$ particles are used in determining $\braket{i|A_k|j}$. Since both the sets $\{\ket{\pm_{\text{P}}}\}$ and $\{\ket{\pm i_{\text{P}}}\}$ are measured, we allocate $\frac{N}{2}$ particles for the measurement of $\{\ket{\pm_{\text{P}}}\}$ and the remaining $\frac{N}{2}$ particles for the measurement of $\{\ket{\pm i_{\text{P}}}\}$. Also let $n_{l_{\text{P}},i_{\text{S}},k_{\text{E}}}$ is the number of particles that have the post-measurement state $\ket{l_{\text{P}}}\in\mathcal{M}_{\text{P}}=\{\ket{\pm_{\text{P}}},\ket{\pm i_{\text{P}}}\}$ when the system and environment are in $\ket{i_{\text{S}}}$ and $\ket{k_{\text{E}}}$, respectively such that the probability is given by
\begin{align}
p(l_{\text{P}},i_{\text{S}},k_{\text{E}})=\frac{n_{l_{\text{P}},i_{\text{S}},k_{\text{E}}}}{N/2}=\frac{2}{N}n_{l_{\text{P}},i_{\text{S}},k_{\text{E}}}.\tag{S3.1-2}\label{S3.1-2}
\end{align}
Now the variance of the measured probability $p(l_{\text{P}},i_{\text{S}},k_{\text{E}})$ is given by
\begin{align}
\delta^2p(l_{\text{P}},i_{\text{S}},k_{\text{E}})=\frac{4}{N^2}\delta^2n_{l_{\text{P}},i_{\text{S}},k_{\text{E}}}=\frac{4}{N^2}n_{l_{\text{P}},i_{\text{S}},k_{\text{E}}}=\frac{2}{N}p(l_{\text{P}},i_{\text{S}},k_{\text{E}}),\tag{S3.1-3}\label{S3.1-3}
\end{align}
where we considered that the statistic follows the Poissonian statistic and hence the variance of $n_{l_{\text{P}},i_{\text{S}}}$ is equal to $n_{l_{\text{P}},i_{\text{S}}}$, and used Eq. \eqref{S3.1-2}. The error in estimating the matrix element $\braket{i|A_k|j}$ defined in Eq. \eqref{S3-1} using Eqs. \eqref{S3.1-1} and \eqref{S3.1-3} is given by
\begin{align}
|\delta\braket{i|A_k|j}|^2&=\frac{2d_{\text{S}}}{\alpha_kN}\sum_{l_{\text{P}}\in\mathcal{M}_{\text{P}}}p(l_{\text{P}},i_{\text{S}},k_{\text{E}})=\frac{4d_{\text{S}}}{\alpha_kN}p(i_{\text{S}},k_{\text{E}}),\tag{S3.1-4}\label{S3.1-4}
\end{align}
where $\sum_{l_{\text{P}}\in\{\ket{\pm_{\text{P}}}\}}p(l_{\text{P}},i_{\text{S}},k_{\text{E}})=p(i_{\text{S}},k_{\text{E}})=\sum_{l_{\text{P}}\in\{\ket{\pm i_{\text{P}}}\}}p(l_{\text{P}},i_{\text{S}},k_{\text{E}})$ is the marginal probability. \par
Now the  error in estimating $E_k$ as given in Eq.~\eqref{S3-3} is:
\begin{align}
\delta A_k&=\sqrt{\sum_{i,j=0}^{d_{\text{S}}-1}|\delta\braket{i|A_k|j}|^2}=\sqrt{\frac{4d_{\text{S}}}{\alpha_kN}\sum_{i,j=0}^{d_{\text{S}}-1}p(i_{\text{S}},k_{\text{E}})}.\tag{S3.1-5}\label{S3.1-5}
\end{align}
Now from Eq. \eqref{S1-1}, it can be  shown for pure state $\rho_{\text{S}}=\ketbra{j}{j}$ [see Eq. \eqref{DFCKO-1}] that
\begin{align}
\sum_{i,j=0}^{d_{\text{S}}-1}p(i_{\text{S}},k_{\text{E}})&=\sum_{i,j=0}^{d_{\text{S}}-1}\left[\frac{1}{4}|\braket{\xi_{\text{E}}|k_{\text{E}}}|^2\braket{i|U_{\text{S}}\ketbra{j}{j}U_{\text{S}}^{\dagger}|i}+\frac{1}{4} \bra{i}\otimes\bra{k_{\text{E}}}\left[U_{\text{SE}}\left(\ketbra{j}{j}\otimes\ketbra{\xi_{\text{E}}}{\xi_{\text{E}}}\right)U_{\text{SE}}^{\dagger}\right]\ket{i}\otimes\ket{k_{\text{E}}}\right]\nonumber\\
&=\frac{1}{4}\left[\alpha_kd_{\text{S}}+Tr(E_k)\right],\tag{S3.1-6}\label{S3.1-6}
\end{align}
where we have used the definition of $A_k$ and $A^{\dagger}_kA_k=E_k$. Now we define the error in estimating $E_k$ as $\delta E_k=\delta A_k$, and by substituting Eq. \eqref{S3.1-6} in Eq. \eqref{S3.1-5}, we have
\begin{align}
\delta E_k=\sqrt{\frac{d_{\text{S}}}{\alpha_kN}\left[\alpha_kd_{\text{S}}+Tr(E_k)\right]}.\tag{S3.1-7}\label{S3.1-7}
\end{align}
The error in estimating the POVM, $E=\{E_k\}_{k=0}^{d_{\text{E}}-1}$ defined in Eq. \eqref{S3-4} is:
\begin{align}
\delta E=\sqrt{\sum_{k=0}^{d_{\text{E}}-1}\delta E_k^2}=\sqrt{\sum_{k=0}^{d_{\text{E}}-1}\frac{d_{\text{S}}}{\alpha_kN}\left[\alpha_kd_{\text{S}}+Tr(E_k)\right]}=d_{\text{S}}\sqrt{\frac{2d_{\text{E}}}{N}},\tag{S3.1-8}\label{S3.1-8}
\end{align}
where we have taken $\alpha_k=\frac{1}{d_{\text{E}}}$ and used $\sum_{k=0}^{d_{\text{E}}-1}E_k=I_{\text{S}}$, and note that $Tr(E_k)\leq d_{\text{S}}$.

\subsection{S3.2. By the method given in Ref. \cite{Xu-2021-DirectMeasurements}}
In Ref. \cite{Xu-2021-DirectMeasurements}, the $(i,j)$-th element of $E_k$ is given by
\begin{align}
\braket{i|E_k|j}=\sum_{l=0}^{d_{\text{S}}-1}e^{2\pi i(i-j)\frac{l}{d_{\text{S}}}}\omega_{j,l}^{(k)},\tag{S3.2-1}\label{S3.2-1}
\end{align}
where 
\begin{align}
\omega_{j,l}^{(k)}=\frac{1}{2\sin\theta}\left[p_{l}^j(+_{\text{P}},k_{\text{S}})-p_{l}^j(-_{\text{P}},k_{\text{S}})+2\tan\frac{\theta}{2} p_{l}^j(1_{\text{P}},k_{\text{S}})+i\{p_{l}^j(+i_{\text{P}},k_{\text{S}})-p_{l}^j(-i_{\text{P}},k_{\text{S}})\}\right],\tag{S3.2-2}\label{S3.2-2}
\end{align}
and 
\begin{align}
p_{l}^j(r_{\text{P}},k_{\text{S}})=Tr\left[(\ketbra{r_{\text{P}}}{r_{\text{P}}}\otimes E_k)e^{-i\theta\sigma_{\text{P}}^y \otimes\ketbra{j_{\text{S}}}{j_{\text{S}}}}(\ketbra{\chi_{\text{P}}}{\chi_{\text{P}}}\otimes\ketbra{l_{\text{S}}}{l_{\text{S}}})e^{i\theta\sigma_{\text{P}}^y \otimes\ketbra{j_{\text{S}}}{j_{\text{S}}}}\right],\tag{S3.2-3}\label{S3.2-3}
\end{align}
where $\ket{r_{\text{P}}}\in\{\ket{\pm_{\text{P}}},\ket{\pm i_{\text{P}}},\ket{1_{\text{P}}}\}$. After substituting Eq. \eqref{S3.2-2} in Eq. \eqref{S3.2-1}, we have the real and imaginary parts:
\begin{align}
\text{Re}[\braket{i|E_k|j}]&=\frac{1}{2\sin\theta}\sum_{l=0}^{d_{\text{S}}-1}\Bigg\{\cos\left(2\pi l\frac{i-j}{d_{\text{S}}}\right)\left[p_{l}^j(+_{\text{P}},k_{\text{S}})-p_{l}^j(-_{\text{P}},k_{\text{S}})+2\tan\frac{\theta}{2} p_{l}^j(1_{\text{P}},k_{\text{S}})\right]\nonumber\\
&\quad\quad\quad\quad\quad\quad\quad-\sin\left(2\pi l\frac{i-j}{d_{\text{S}}}\right)\left[p_{l}^j(+i_{\text{P}},k_{\text{S}})-p_{l}^j(-i_{\text{P}},k_{\text{S}})\right]\Bigg\}.\tag{S3.2-4}\label{S3.2-4}
\end{align}
\begin{align}
\text{Im}[\braket{i|E_k|j}]&=\frac{1}{2\sin\theta}\sum_{l=0}^{d_{\text{S}}-1}\Bigg\{\cos\left(2\pi l\frac{i-j}{d_{\text{S}}}\right)\left[p_{l}^j(+i_{\text{P}},k_{\text{S}})-p_{l}^j(-i_{\text{P}},k_{\text{S}})\right]   \nonumber\\
&\quad\quad\quad\quad\quad\quad\quad-\sin\left(2\pi l\frac{i-j}{d_{\text{S}}}\right)\left[p_{l}^j(+_{\text{P}},k_{\text{S}})-p_{l}^j(-_{\text{P}},k_{\text{S}})+2\tan\frac{\theta}{2} p_{l}^j(1_{\text{P}},k_{\text{S}})\right]\Bigg\}.\tag{S3.2-5}\label{S3.2-5}
\end{align}
Now we assume that $N$ particles are used in determining $\braket{i|E_k|j}$. Since three sets $\{\ket{\pm_{\text{P}}}\}$, $\{\ket{\pm i_{\text{P}}}\}$, and $\{\ket{0_{\text{P}}},\ket{1_{\text{P}}}\}$ are measured, we allocate $\frac{N}{3}$ particles for the measurement of $\{\ket{\pm_{\text{P}}}\}$, $\frac{N}{3}$ particles for the measurement of $\{\ket{\pm i_{\text{P}}}\}$ and the remaining $\frac{N}{3}$ particles for the measurement of $\{\ket{0_{\text{P}}},\ket{1_{\text{P}}}\}$. Also let $n^{l,j}_{r_{\text{P}},k_{\text{S}}}$ is the number of particles that have the post-measurement state $\ket{r_{\text{P}}}\in\mathcal{M}_{\text{P}}=\{\ket{\pm_{\text{P}}},\ket{\pm i_{\text{P}}}, \ket{1_{\text{P}}}\}$ when the probe and system are in $\ket{r_{\text{P}}}$ and $\ket{k_{\text{S}}}$, respectively when the initial state of the system and the unitary evolution of the probe-system are $\ket{l_{\text{S}}}$, and $e^{i\theta\sigma_{\text{P}}^y\otimes\ketbra{j_{\text{S}}}{j_{\text{S}}}}$  such that the probability is given by
\begin{align}
p_l^j(r_{\text{P}},k_{\text{S}})=\frac{n^{l,j}_{r_{\text{P}},k_{\text{S}}}}{N/3}=\frac{3}{N}n^{l,j}_{r_{\text{P}},k_{\text{S}}}.\tag{S3.2-6}\label{S3.2-6}
\end{align}
Now the variance of the measured probability $p^j_l(r_{\text{P}},k_{\text{S}})$ is given by
\begin{align}
\delta^2p^j_l(r_{\text{P}},k_{\text{S}})=\frac{9}{N^2}\delta^2n^{l,j}_{r_{\text{P}},k_{\text{S}}}=\frac{9}{N^2}n^{l,j}_{r_{\text{P}},k_{\text{S}}}=\frac{3}{N}p^j_l(r_{\text{P}},k_{\text{S}}),\tag{S3.2-7}\label{S3.2-7}
\end{align}
where we considered that the statistic follows the Poissonian statistic and hence the variance of $n^{l,j}_{r_{\text{P}},k_{\text{S}}}$ is equal to $n^{l,j}_{r_{\text{P}},k_{\text{S}}}$, and used Eq. \eqref{S3.2-6}.\par
\begin{figure}[H]
\centering
\includegraphics[scale=0.53]{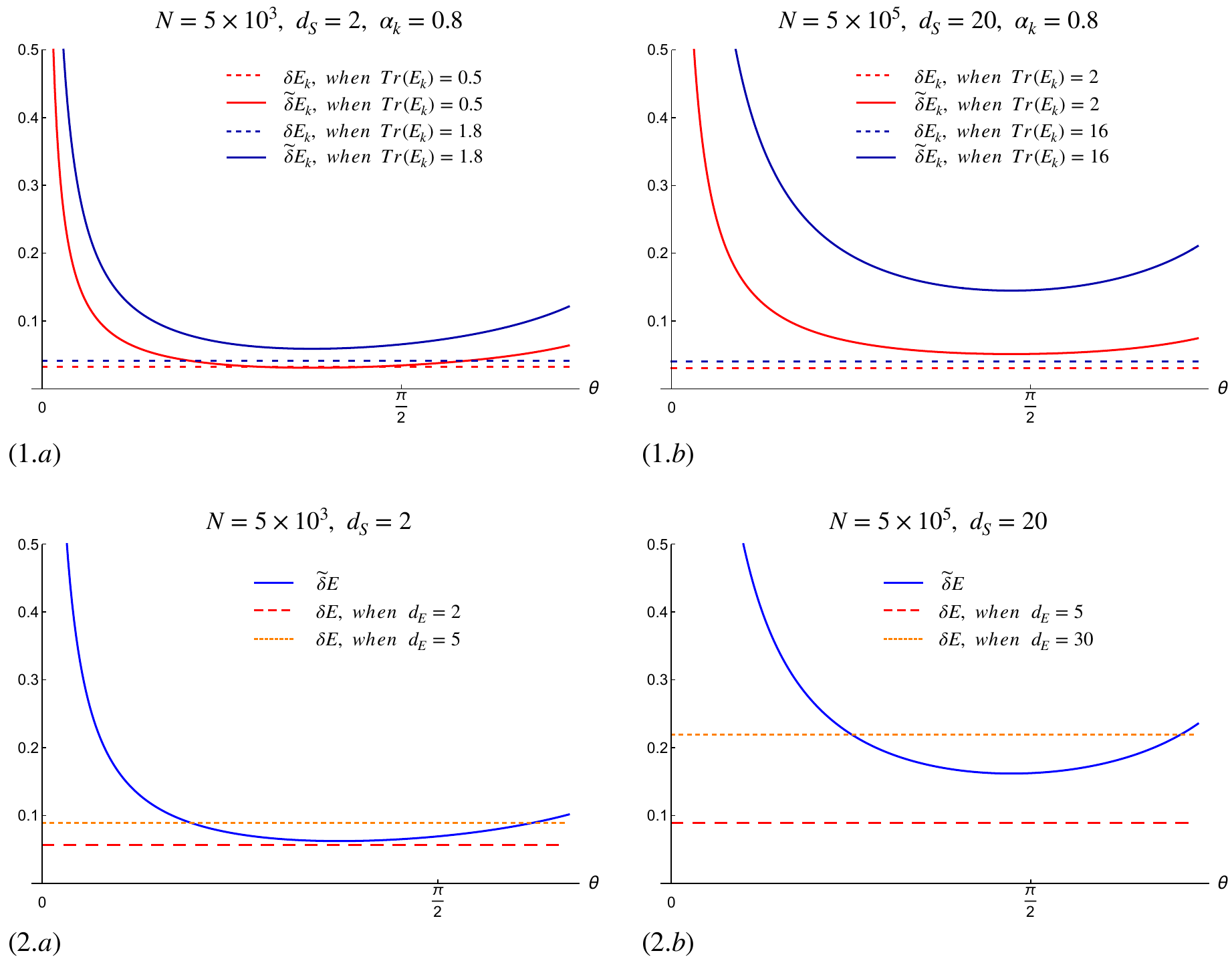}
\caption{
Comparison of our results in Eqs.~\eqref{S3.1-7} and \eqref{S3.1-8} with those of Ref.~\cite{Xu-2021-DirectMeasurements} for the error in estimating an unknown POVM element $E_k$ as well as the entire POVM, $E = \{E_k\}_{k=0}^{d_{\text{E}}-1}$. In (1.a), we plot $\delta E_k$ [our result from Eq.~\eqref{S3.1-7}] versus $\widetilde{\delta}E_k$ [the result of Ref.~\cite{Xu-2021-DirectMeasurements}, recalculated in Eq.~\eqref{S3.2-11}]. For $d_{\text{S}} = 2$, if the POVM element $E_k$ has a low trace value, \emph{e.g.,} $\text{Tr}(E_k) = 0.5$, both methods perform comparably. However, for high trace values, \emph{e.g.,} $\text{Tr}(E_k) = 1.8$, our method significantly outperforms that of Ref.~\cite{Xu-2021-DirectMeasurements} for any value of the unitary interaction parameter $\theta$ introduced therein. A similar performance trend is observed in (1.b) for $d_{\text{S}} = 20$, indicating that the performance gain of our method increases with system dimension. In (2.a), we compare $\delta E$ [our result from Eq.~\eqref{S3.1-8}] with $\widetilde{\delta}E$ [from Ref.~\cite{Xu-2021-DirectMeasurements}, recalculated in Eq.~\eqref{S3.2-12}]. For $d_{\text{S}} = 2$ and $d_{\text{E}} = 2$, our method performs slightly better. However, when $d_{\text{E}} \geq d_{\text{S}}$, \emph{e.g.,} $d_{\text{E}} = 5$, the method of Ref.~\cite{Xu-2021-DirectMeasurements} performs better. In (2.b), for $d_{\text{S}} = 20$, a similar trend holds, but when the difference $d_{\text{S}} - d_{\text{E}} \gg 1$ (\emph{e.g.,} $d_{\text{S}} - d_{\text{E}} = 15$), our method again outperforms the method of Ref. \cite{Xu-2021-DirectMeasurements}.}
\label{FIG3}
\end{figure}

The error in estimating $E_k$ as given in Eq. \eqref{S3-3} is: 
\begin{align}
\widetilde{\delta} E_k^2&=\sum_{i,j=0}^{d_{\text{S}}-1}|\delta\braket{i|E_k|j}|^2\nonumber\\
&=\sum_{i,j=0}^{d_{\text{S}}-1}\left(\delta \text{Re}[\braket{i|E_k|j}]^2+\delta \text{Re}[\braket{i|E_k|j}]^2\right)
\nonumber\\
&=\sum_{i,j=0}^{d_{\text{S}}-1}\sum_{l=0}^{d_{\text{S}}-1}\sum_{r\in \mathcal{M}_{\text{P}}} \left[\left(\frac{\partial \text{Re}[\braket{i|E_k|j}]}{\partial p^j_l(r_{\text{P}},k_{\text{S}})}\right)^2\delta^2p_l^j(r_{\text{P}},k_{\text{S}})+\left(\frac{\partial \text{Im}[\braket{i|E_k|j}]}{\partial p^j_l(r_{\text{P}},k_{\text{S}})}\right)^2\delta^2p_l^j(r_{\text{P}},k_{\text{S}})\right],\nonumber\\
&=\frac{1}{4\sin^2\theta}\frac{3}{N}\sum_{i,j=0}^{d_{\text{S}}-1}\sum_{l=0}^{d_{\text{S}}-1}\left[\{2p_l^j(k_{\text{S}})+4\tan^2\frac{\theta}{2}p_{l}^j(1_{\text{P}},k_{\text{S}})\right].\tag{S3.2-8}\label{S3.2-8}
\end{align}
Now let us evaluate the following using Eq. \eqref{S3.2-3}:
\begin{align}
\sum_{i,j=0}^{d_{\text{S}}-1}\sum_{l=0}^{d_{\text{S}}-1}p_l^j(k_{\text{S}})&=\sum_{i,j=0}^{d_{\text{S}}-1}\sum_{l=0}^{d_{\text{S}}-1}Tr\left[(I_{\text{P}}\otimes E_k)e^{-i\theta\sigma_{\text{P}}^y \otimes\ketbra{j_{\text{S}}}{j_{\text{S}}}}(\ketbra{\chi_{\text{P}}}{\chi_{\text{P}}}\otimes\ketbra{l_{\text{S}}}{l_{\text{S}}})e^{i\theta\sigma_{\text{P}}^y \otimes\ketbra{j_{\text{S}}}{j_{\text{S}}}}\right],\nonumber\\
&=\sum_{i,j=0}^{d_{\text{S}}-1}Tr\left[(I_{\text{P}}\otimes E_k)e^{-i\theta\sigma_{\text{P}}^y \otimes\ketbra{j_{\text{S}}}{j_{\text{S}}}}(\ketbra{\chi_{\text{P}}}{\chi_{\text{P}}}\otimes I_{\text{S}})e^{i\theta\sigma_{\text{P}}^y \otimes\ketbra{j_{\text{S}}}{j_{\text{S}}}}\right]=d_{\text{S}}^2Tr(E_k),\tag{S3.2-9}\label{S3.2-9}
\end{align}
where we used the identities: $\sum_{l=0}^{d_{\text{S}}-1}\ketbra{l_{\text{S}}}{l_{\text{S}}}=I_{\text{S}}$ and $e^{\pm i\theta\sigma_{\text{P}}^y \otimes\ketbra{j_{\text{S}}}{j_{\text{S}}}}=I_{\text{P}}\otimes (I_{\text{S}}-\ketbra{j_{\text{S}}}{j_{\text{S}}})+e^{\pm i\theta\sigma_{\text{P}}^y}\otimes\ketbra{j_{\text{S}}}{j_{\text{S}}}$.
Also evaluate the following using Eq. \eqref{S3.2-3}:
\begin{align}
\sum_{i,j=0}^{d_{\text{S}}-1}\sum_{l=0}^{d_{\text{S}}-1}p_{l}^j(1_{\text{P}},k_{\text{S}})&=\sum_{i,l=0}^{d_{\text{S}}-1}\frac{1}{d_{\text{S}}}\sin^2\theta Tr(E_k)=d_{\text{S}}\sin^2\theta Tr(E_k),\tag{S3.2-10}\label{S3.2-10}
\end{align}
where we have taken $\ket{\chi_{\text{P}}}=\ket{0_{\text{P}}}$. By using  Eqs. \eqref{S3.2-9} and \eqref{S3.2-10} in \eqref{S3.2-8}, we have
\begin{align}
\widetilde{\delta} E_k^2&=\frac{6d_{\text{S}}}{4N\sin^2\theta}Tr(E_k)\left[d_{\text{S}}+2\sin^2\theta\tan^2\frac{\theta}{2} \right].\tag{S3.2-11}\label{S3.2-11}
\end{align}
The error in estimating the POVM, $E=\{E_k\}_{k=0}^{d_{\text{E}}-1}$ defined in Eq. \eqref{S3-4} is:
\begin{align}
\widetilde{\delta} E=\sqrt{\sum_{k=0}^{d_{\text{E}}-1}\widetilde{\delta} E_k^2}=\frac{d_{\text{S}}}{\sin\theta}\sqrt{\frac{3}{2N}\left[d_{\text{S}}+2\sin^2\theta\tan^2\frac{\theta}{2} \right]}.\tag{S3.2-12}\label{S3.2-12}
\end{align}
In Fig.~\ref{FIG3}, we compare our results, Eqs.~\eqref{S3.1-7} and \eqref{S3.1-8}, with those recalculated from Ref.~\cite{Xu-2021-DirectMeasurements}, as shown in Eqs.~\eqref{S3.2-11} and \eqref{S3.2-12}. Under certain restricted conditions, the method of Ref.~\cite{Xu-2021-DirectMeasurements}, represented by Eqs.~\eqref{S3.2-11} and \eqref{S3.2-12}, can outperform our approach. However, as discussed in the caption of Fig.~\ref{FIG3}, their method becomes significantly inaccurate when the interaction coupling $\theta$ between the probe and system approaches zero. Additional limitations of their approach are addressed in the main text.

\section{S4. Characterization of a Kraus operator by a pure input state and d-dimensional Pauli X-gates}
Here, we show that a single pure state, namely $\ket{0}$, prepared initially in the system, is sufficient to characterize the Kraus operator $A_k$ using the Pauli $X$-gate and its higher-order generalizations~\cite{Babazadeh-2017,Gao-Zeilinger-2019,Wang_2022,Meng-2024}. The $n$-th order Pauli-$X$ gate is defined as $X^n_{\text{S}}\ket{i} = \ket{(i+n)\,\text{modulo}\,d_{\text{S}}}$, while its adjoint acts as ${X^n_{\text{S}}}^{\dagger}\ket{i} = \ket{(i-n)\,\text{modulo}\,d_{\text{S}}} = \ket{j}$.\par
To obtain the $(i,j)$-th element of the Kraus operator $A_k$, we
set:
\begin{itemize}
\item $\rho_S=\ketbra{0}{0}$,
\item $\ket{\phi_S}=\ket{i}$,
\item $U_S$ to be such that $\braket{0|U_S^{\dagger}|i}\neq 0$ $\,\,\,\forall\,\,i$,
\item $\widetilde{U}_S\ket{0}=\ket{j}$
\end{itemize}
in Eq.~(\ref{FW-3}), which yields the following:
\begin{align}
\braket{i|A_k|j}=d_S\left[\frac{\braket{ (\sigma^x_{\text{P}} + i \sigma^y_{\text{P}}) \otimes \Pi^{i}_{\text{S}} \otimes \Pi^{k}_{\text{E}} }_{\rho(t)}}{\mathcal{N}_{\text{PSE}}\braket{0|U_S^{\dagger}|i}}\right].\tag{S4-1}\label{S4-1}
\end{align} 
For the full characterization, take $\widetilde{U}_S$ to be the $j$-th order Pauli X-gate \emph{i.e.,} $\widetilde{U}_S=X^j_S$ whose action is defined as: $X^j_S\ket{0}=\ket{j}$. By considering $j=0,1,\cdots,d_S-1$, we achieve the whole matrix form of $A_k$.\par
Since the resources required to fully characterize the Kraus operator $A_k$ are the same as the method given in Eq.~\eqref{DFCKO-1}, the error in estimating the corresponding POVM element $E_k = A_k^{\dagger} A_k$ using the method in Eq.~\eqref{S4-1} can be shown to be equivalent to that of the method in Eq.~\eqref{DFCKO-1}. See Sec.~S3.1 of the Supplemental Material for a detailed discussion on the error estimation of $A_k$ using the method in Eq.~\eqref{DFCKO-1}.

\section{S5. Direct characterization of an unknown \emph{density matrix} in a modified Mach–Zehnder interferometer (MZI) using our method}
Although our method of direct characterization of density matrix works in any situation described in Eq. (\ref{DFCDM-1}), here we propose a slightly different situation where the unknown density matrix of our interest is prepared by letting it interact with the environment and tracing out the environment during the time evolution inside the modified MZI.
\begin{figure}[H]
\centering
\includegraphics[scale=0.53]{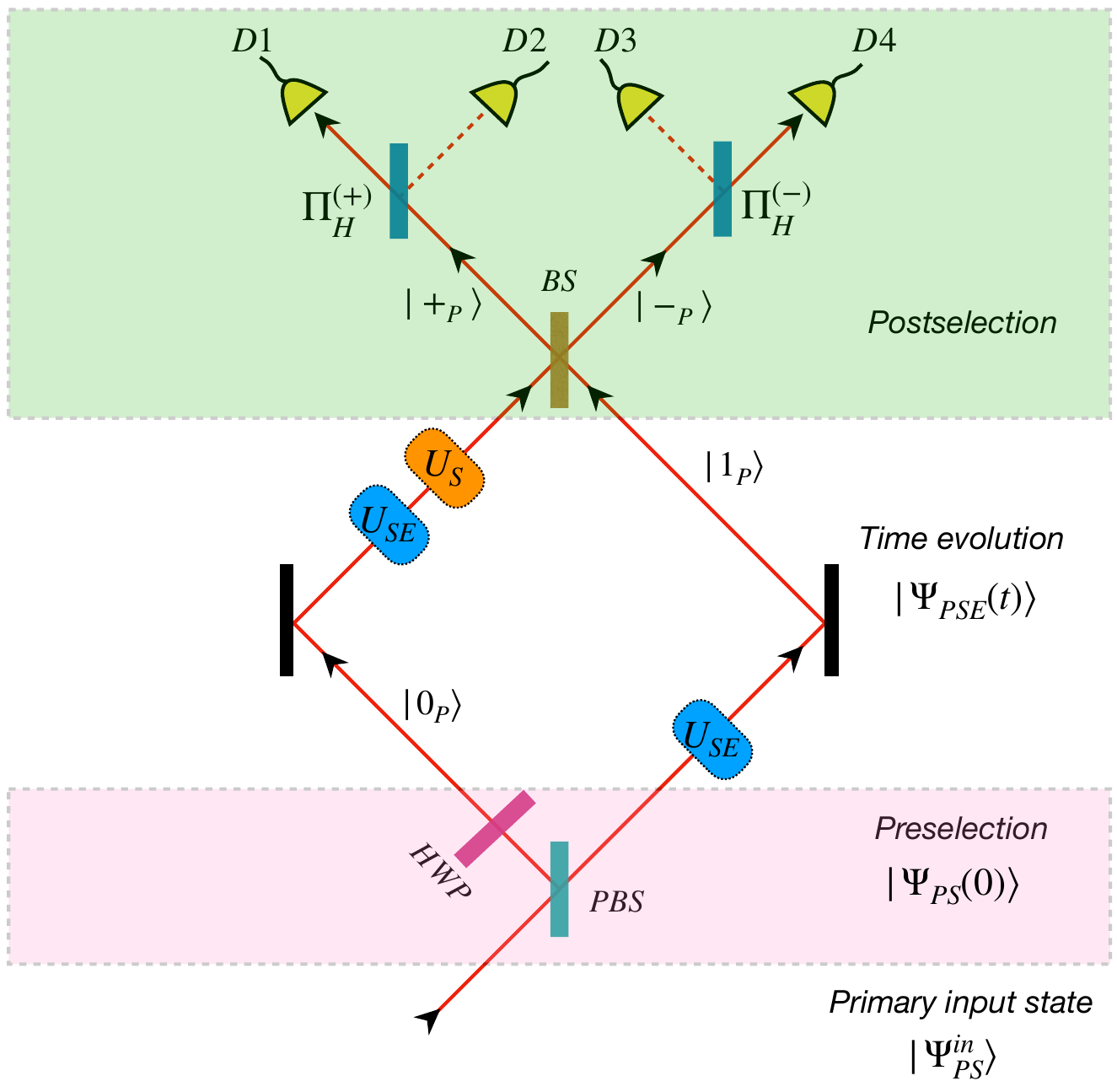}
\caption{Schematic illustration of a modified Mach–Zehnder interferometer (MZI) used for the direct characterization of an unknown density matrix.}
\label{FIG4}
\end{figure}
To realize this, we consider an experimental setup as depicted in Fig. \ref{FIG4}, which is a modified version of the standard Mach–Zehnder interferometer. In this arrangement-similar to the previous one-a single photon is used, where the photon's path degree of freedom serves as the probe, and its polarization degree of freedom represents the system. The environment is considered as spin-$1/2$ particles, \emph{i.e.,} qubits.\par
The basis states for the probe, system, and environment are denoted as $\{\ket{0_{\text{P}}}, \ket{1_{\text{P}}}\}$, $\{\ket{H}, \ket{V}\}$, and $\{\ket{0_{\text{E}}}, \ket{1_{\text{E}}}\}$, respectively. Our goal is to determine the matrix element $\braket{H|\rho|V}$, where the density matrix $\rho$ is considered to be a transformed one:\\
\begin{align*}
\ketbra{V}{V}\otimes\ketbra{\xi_{\text{E}}}{\xi_{\text{E}}}\xrightarrow{Tr_{\text{E}}\circ U_{\text{SE}}}\rho=K_0\ketbra{V}{V}K^{\dagger}_0+K_1\ketbra{V}{V}K^{\dagger}_1,
\end{align*}
where the environment is initially in a pure uncorrelated state $\ket{\xi_{\text{E}}}$. The Kraus operators are: $A_0 = \braket{0_{\text{E}}|U_{\text{SE}}|\xi_{\text{E}}}$ and $A_1 = \braket{1_{\text{E}}|U_{\text{SE}}|\xi_{\text{E}}}$ defined in Eq.~(\ref{FW-3}), during the time evolution inside the MZI; see Fig. \ref{FIG4}.\par
We begin by preparing the joint initial state:
\begin{align*}
\ket{\Psi_{\text{PS}}(0)}\otimes \ket{\xi_{\text{E}}} = (\cos\alpha\ket{0_{\text{P}}} + \sin\alpha\ket{1_{\text{P}}}) \otimes \ket{V} \otimes \ket{\xi_{\text{E}}}.
\end{align*}
To achieve this, we start with the photon's primary input state $\ket{\Psi^{\text{in}}_{\text{PS}}} = \ket{a_{\text{P}}} \otimes (\cos\alpha\ket{H} + \sin\alpha\ket{V})$. Upon passing through a polarizing beam splitter (PBS), the state transforms as:
\begin{align*}
\ket{\Psi^{\text{in}}_{\text{PS}}} \xrightarrow{PBS} \cos\alpha \ket{0_{\text{P}}} \otimes \ket{H} + \sin\alpha \ket{1_{\text{P}}} \otimes \ket{V},
\end{align*}
creating entanglement between path (probe) and polarization (system). Next, a half-wave plate (HWP) is placed in path $\ket{0_{\text{P}}}$ to flip polarization: $\ket{H} \xrightarrow{HWP} \ket{V}$, resulting in the desired preselected state of the probe-system,
\begin{align*}
\ket{\Psi_{\text{PS}}(0)} = (\cos\alpha\ket{0_{\text{P}}} + \sin\alpha\ket{1_{\text{P}}}) \otimes \ket{V}.
\end{align*}
Before entering the evolution region of the modified MZI, the total state is $\ket{\Psi_{\text{PS}}(0)}\otimes\ket{\xi_{\text{E}}}$. Upon entering --- if the photon follows path $\ket{0_{\text{P}}}$ --- it evolves as:
\begin{align*}
\ket{0_{\text{P}}} \otimes\ket{V}\otimes \ket{\xi_{\text{E}}} \xrightarrow{I_{\text{P}} \otimes U_{\text{SE}}} \ket{0_{\text{P}}} \otimes U_{\text{SE}} \ket{V}\otimes \ket{\xi_{\text{E}}} \xrightarrow{I_{\text{P}} \otimes U_{\text{S}}\otimes I_{\text{E}}} \ket{0_{\text{P}}}\otimes U_{\text{S}}U_{\text{SE}} \ket{V} \otimes\ket{\xi_{\text{E}}},
\end{align*}
if it takes path $\ket{1_{\text{P}}}$, it undergoes:
\begin{align*}
\ket{1_{\text{P}}}\otimes \ket{V}\otimes \ket{\xi_{\text{E}}} \xrightarrow{I_{\text{P}} \otimes U_{\text{SE}}} \ket{1_{\text{P}}}\otimes U_{\text{SE}} \ket{V} \otimes\ket{\xi_{\text{E}}}.
\end{align*}
Thus, after the time evolution, the total state becomes:
\begin{align*}
\ket{\Psi_{\text{PSE}}(t)} = \cos\alpha \ket{0_{\text{P}}}\otimes U_{\text{S}}U_{\text{SE}} \ket{V}\otimes \ket{\xi_{\text{E}}} + \sin\alpha \ket{1_{\text{P}}}\otimes U_{\text{SE}} \ket{V}\otimes \ket{\xi_{\text{E}}}.
\end{align*}
Now the environment is measured in the state $\ket{0_{\text{E}}}$, yielding the probe-system state proportional to:
\begin{align*}
\cos\alpha \ket{0_{\text{P}}}\otimes U_{\text{S}}A_0\ket{V} + \sin\alpha \ket{1_{\text{P}}}\otimes A_0\ket{V}, 
\end{align*}
where $A_0 = \braket{0_{\text{E}}|U_{\text{SE}}|\xi_{\text{E}}}$ is the kraus operator when the operator $\Pi_{0_{\text{E}}}=\ketbra{0_{\text{E}}}{0_{\text{E}}}$ is measured on the environment.\par
The beam splitter (BS) is defined by:
\begin{align*}
\ket{0_{\text{P}}} \xrightarrow{BS} \frac{1}{\sqrt{2}}(\ket{+_{\text{P}}} + \ket{-_{\text{P}}}), \quad
\ket{1_{\text{P}}} \xrightarrow{BS} \frac{1}{\sqrt{2}}(\ket{+_{\text{P}}} - \ket{-_{\text{P}}}),
\end{align*}
so that the unnormalized probe-system state after the BS becomes:
\begin{align*}
\frac{1}{\sqrt{2}} \ket{+_{\text{P}}}\otimes[\cos\alpha\, U_{\text{S}}A_0\ket{V} + \sin\alpha\, A_0\ket{V}] + \frac{1}{\sqrt{2}} \ket{-_{\text{P}}}\otimes[\cos\alpha\, U_{\text{S}}A_0\ket{V} - \sin\alpha\, A_0\ket{V}].
\end{align*}
We define the postselection operator $\Pi_H^{(+)} = \ketbra{H}{H}$, which transmits only the horizontal polarization. Hence, if detector $D1$ clicks, we know that the photon is in the state $\ket{+_{\text{P}}} \otimes \ket{H}\otimes\ket{0_{\text{E}}}$. The probability of this joint detection (\emph{i.e.,} the probe, system, and environment being in the states $\ket{+_{\text{P}}}$, $\ket{H}$, and $\ket{0_{\text{E}}}$, respectively) is:
\begin{align}
p(+_{\text{P}}, H, 0_{\text{E}}) &= \frac{1}{2} \cos^2\alpha|\braket{H|U_{\text{S}}A_0|V}|^2 + \frac{1}{2} \sin^2\alpha\, |\braket{H|A_0|V}|^2 \notag \\
&\quad + \text{Re}\big[\braket{V|A_0^{\dagger}U_{\text{S}}^\dagger|H} \braket{H|A_0|V}\big] \cos\alpha \sin\alpha. \tag{S5-1}\label{S5-1}
\end{align}
Similarly, the probability of joint detection of the probe, system, and environment being in the states $\ket{+_{\text{P}}}$, $\ket{H}$, and $\ket{0_{\text{E}}}$, respectively is:
\begin{align}
p(-_{\text{P}}, H, 0_{\text{E}}) &= \frac{1}{2} \cos^2\alpha|\braket{H|U_{\text{S}}A_0|V}|^2 + \frac{1}{2} \sin^2\alpha\, |\braket{H|A_0|V}|^2 \notag \\
&\quad - \text{Re}\big[\braket{V|A_0^{\dagger}U_{\text{S}}^\dagger|H} \braket{H|A_0|V}\big] \cos\alpha \sin\alpha. \tag{S5-2}\label{S5-2}
\end{align}
Next, we replace the BS with a modified one, denoted by $\widetilde{BS}$, whose action is:
\begin{align*}
\ket{0_{\text{P}}} \xrightarrow{\widetilde{BS}} \frac{1}{\sqrt{2}}(\ket{+i_{\text{P}}} + i \ket{-i_{\text{P}}}), \quad
\ket{1_{\text{P}}} \xrightarrow{\widetilde{BS}} \frac{1}{\sqrt{2}}(\ket{+i_{\text{P}}} - i \ket{-i_{\text{P}}}).
\end{align*}
Here, we simply rename the paths by replacing $\ket{\pm_{\text{P}}}$ with $\ket{\pm i_{\text{P}}}$.\par
Under this transformation, the probabilities of joint detection of the probe, system, and environment being in the states $\ket{\pm i_{\text{P}}}$, $\ket{H}$, and $\ket{0_{\text{E}}}$, respectively is:
\begin{align}
p(+i_{\text{P}}, H, 0_{\text{E}}) &= \frac{1}{2} \cos^2\alpha|\braket{H|U_{\text{S}}A_0|V}|^2 + \frac{1}{2} \sin^2\alpha\, |\braket{H|A_0|V}|^2 \notag \\
&\quad - \text{Im}\big[\braket{V|A_0^{\dagger}U_{\text{S}}^\dagger|H} \braket{H|A_0|V}\big] \cos\alpha \sin\alpha, \tag{S5-3}\label{S5-3}
\end{align}
\begin{align}
p(-i_{\text{P}}, H, 0_{\text{E}}) &= \frac{1}{2} \cos^2\alpha|\braket{H|U_{\text{S}}A_0|V}|^2 + \frac{1}{2} \sin^2\alpha\, |\braket{H|A_0|V}|^2 \notag \\
&\quad - \text{Im}\big[\braket{V|A_0^{\dagger}U_{\text{S}}^\dagger|H} \braket{H|A_0|V}\big] \cos\alpha \sin\alpha. \tag{S5-4}\label{S5-4}
\end{align}
Combining Eqs. (\ref{S5-1})–(\ref{S5-4}), we have the following:
\begin{align*}
\braket{H|A_0|V}\braket{V|A_0^{\dagger}U_{\text{S}}^\dagger|H} &= \frac{p(+_{\text{P}}, H, 0_{\text{E}}) - p(-_{\text{P}}, H, 0_{\text{E}})+ i \big[p(+i_{\text{P}}, H, 0_{\text{E}}) - p(-i_{\text{P}}, H, 0_{\text{E}})\big]}{2 \cos\alpha \sin\alpha}.\tag{S5-5}\label{S5-5}
\end{align*}
In the similar manner, by obtaining the probabilities $p(+_{\text{P}}, H, 1_{\text{E}})$, $p(-_{\text{P}}, H, 1_{\text{E}})$, $p(+i_{\text{P}}, H, 1_{\text{E}})$, and $p(-i_{\text{P}}, H, 1_{\text{E}})$, we have the following:
\begin{align*}
\braket{H|A_1|V}\braket{V|A_1^{\dagger}U_{\text{S}}^\dagger|H} &= \frac{p(+_{\text{P}}, H, 1_{\text{E}}) - p(-_{\text{P}}, H, 1_{\text{E}})+ i \big[p(+i_{\text{P}}, H, 1_{\text{E}}) - p(-i_{\text{P}}, H, 1_{\text{E}})\big]}{2 \cos\alpha \sin\alpha}.\tag{S5-6}\label{S5-6}
\end{align*}
Now, choosing $U_{\text{S}}=\sigma_{\text{S}}^x$ to a Pauli X operator such that $\sigma_{\text{S}}^x\ket{H}=\ket{V}$. Finally we can now experimentally determine the matrix element $\braket{H|\rho|V}$ of the density matrix $\rho$ as:
\begin{align}
\braket{H|\rho|V}=\braket{H|A_0|V}\braket{V|A_0^{\dagger}|V}+\braket{H|A_1|V}\braket{V|A_1^{\dagger}|V},\tag{S5-7}\label{S5-7}
\end{align}
where each of the term in the right-hand side of Eq. (\ref{S5-7}) can experimentally be obtained from Eqs. (\ref{S5-5}) and (\ref{S5-6}).

\section{S6. Full characterization of an unknown \emph{density matrix} using our method given in Eq. \eqref{DFCDM-1} by $d_\text{S}$-dimensional Pauli X-gates}\label{S6}
The $(i,j)$-th matrix element of the density operator $\rho_\text{S}$ is given from Eq. \eqref{DFCDM-1} as
\begin{align}
\braket{i | \rho_\text{S} | j} = \frac{1}{\mathcal{N}_\text{PS}} \braket{ (\sigma^x_\text{P} + i \sigma^y_\text{P}) \otimes \Pi^i_\text{S} }_{\rho(t)}, \tag{S6-1}\label{S6-1}
\end{align}
where $U_\text{S}^\dagger \ket{i} = \ket{j}$.\par
To fully characterize the density matrix $\rho_{\text{S}}$, we use Eq. (\ref{S6-1}) as follows.
We take $U_{\text{S}}$ to be an $n$-th order Pauli X-gate defined as: $X^n_{\text{S}}\ket{i}=\ket{(i+n)\,\,\,modulo\,\,\,d_{\text{S}}}$ while ${X^n_{\text{S}}}^{\dagger}\ket{i}=\ket{(i-n)\,\,\,modulo\,\,\,d_{\text{S}}}=\ket{j}$.
\begin{itemize}
\item  If $U_{\text{S}}=X^0_{\text{S}}=I_{\text{S}}$ is an identity operator, then for $\{\ket{i}\}=\{\ket{0},\ket{1},\cdots,\ket{d_{\text{S}}-1}\}$, we have the matrix elements:
\begin{align*}
\{\braket{i|\rho_{\text{S}}|i\,\,\,modulo\,\,\,d_{\text{S}}}\}_{i=0}^{d_{\text{S}}-1},
\end{align*}
that is the diagonal elements. 
\end{itemize}
\begin{itemize}
\item  If $U_{\text{S}}=X^1_{\text{S}}$, then for $\{\ket{i}\}=\{\ket{0},\ket{1},\cdots,\ket{d_{\text{S}}-1}\}$, we have the matrix elements:
\begin{align*}
\{\braket{i|\rho_{\text{S}}|(i-1)\,\,\,modulo\,\,\,d_{\text{S}}}\}_{i=0}^{d_{\text{S}}-1}.
\end{align*}
\item  If we consider $U_{\text{S}}=X^2_{\text{S}}$, then for $\{\ket{i}\}=\{\ket{0},\ket{1},\cdots,\ket{d_{\text{S}}-1}\}$, we have the matrix elements:
\begin{align*}
\{\braket{i|\rho_{\text{S}}|(i-2)\,\,\,modulo\,\,\,d_{\text{S}}}\}_{i=0}^{d_{\text{S}}-1}.
\end{align*}
\item 
\item 
\item 
\item Finally, if $d_{\text{S}}$ is even, then we consider $U_{\text{S}}=X^{\frac{d_{\text{S}}}{2}}_{\text{S}}$, and for $\{\ket{i}\}=\{\ket{0},\ket{1},\cdots,\ket{d_{\text{S}}-1}\}$, we have the matrix elements:
\begin{align*}
\{\braket{i|\rho_{\text{S}}|(i-\frac{d_{\text{S}}}{2})\,\,\,modulo\,\,\,d_{\text{S}}}\}_{i=0}^{d_{\text{S}}-1}.
\end{align*}
\item  If $d_{\text{S}}$ is odd, then we consider $U_{\text{S}}=X^{\frac{d_{\text{S}}-1}{2}}_{\text{S}}$, and for $\{\ket{i}\}=\{\ket{0},\ket{1},\cdots,\ket{d_{\text{S}}-1}\}$, we have the matrix elements:
\begin{align*}
\{\braket{i|\rho_{\text{S}}|(i-\frac{d_{\text{S}}-1}{2})\,\,\,modulo\,\,\,d_{\text{S}}}\}_{i=0}^{d_{\text{S}}-1}.
\end{align*}
\end{itemize}
This procedure is illustrated with an example given in the following for $d_\text{S}=10$.\par
The number of required Pauli \( X \)-gate and its higher-orders is given by:
 \[
   Pauli\,\, X\!-\!gates=\!
    \begin{cases}
   X^0_{\text{S}},X^1_{\text{S}}, X^2_{\text{S}},\cdots,X^{\frac{d_{\text{S}}}{2}}_{\text{S}} \,\,\,\,\,\,\, \text{if $d_{\text{S}}$ is even \emph{i.e.,}}\quad \frac{d_{\text{S}}}{2}+1,\\
    X^0_{\text{S}},X^1_{\text{S}}, X^2_{\text{S}},\cdots,X^{\frac{d_{\text{S}}-1}{2}}_{\text{S}} \,\,\,\,\,\,\text{if $d_{\text{S}}$ is odd \emph{i.e.,}}\quad \frac{d_{\text{S}}-1}{2}+1,
    \end{cases}
\]
for full characterization of an unknown density matrix of dimension $d_{\text{S}}$.

\begin{figure}[H]
\centering
\includegraphics[scale=0.5]{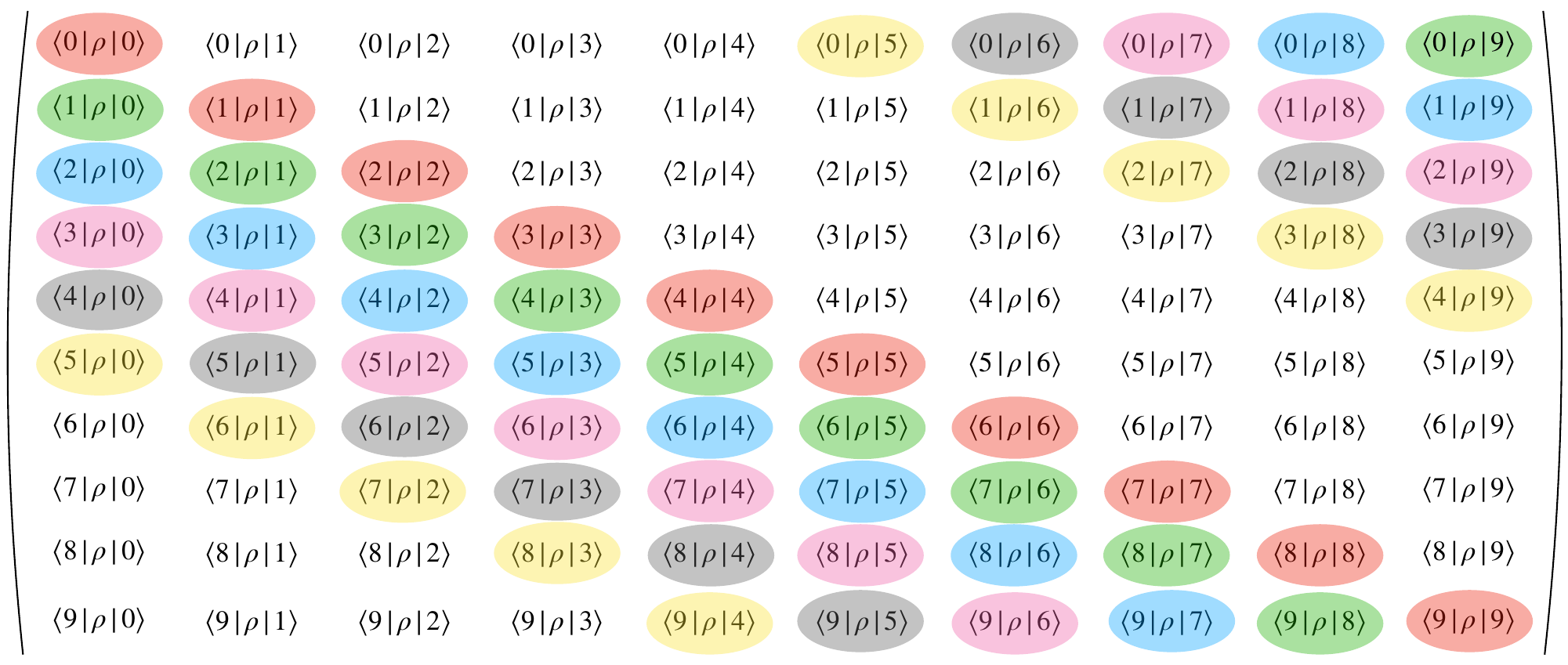}
\caption{Description of this figure is given in the ``Example: $d_\text{S}=10$".}
\label{FIG5}
\end{figure}

\subsection*{Example: $d_\text{S}=10$}
We consider the dimension of the system to be nine \emph{i.e.,} $d_{\text{S}}=10$ and $\{\ket{\phi_{\text{S}}}\}=\{\ket{0},\ket{1},\cdots,\ket{9}\}$. The $n$-th order Pauli X-gate defined as: $X^n_{\text{S}}\ket{i}=\ket{(i+n)\,\,\,modulo\,\,\,10}$ while ${X^n_{\text{S}}}^{\dagger}\ket{i}=\ket{(i-n)\,\,\,modulo\,\,\,10}$, where $i,n=0,1,\cdots,9$.
\begin{itemize}
\item If $U_{\text{S}}=X^0_{\text{S}}=I_{\text{S}}$, we have
\begin{align*}
\braket{0|\rho_{\text{S}}|0},\braket{1|\rho_{\text{S}}|1},\braket{2|\rho_{\text{S}}|2},\braket{3|\rho_{\text{S}}|3},\braket{4|\rho_{\text{S}}|4},\braket{5|\rho_{\text{S}}|5},\braket{6|\rho_{\text{S}}|6},\braket{7|\rho_{\text{S}}|7},\braket{8|\rho_{\text{S}}|8},\braket{9|\rho_{\text{S}}|9}.
\end{align*}
Refer to Fig.~\ref{FIG5}, where the elements are shown in red within outlined regions.
\item If $U_{\text{S}}=X^1_{\text{S}}$, we have
\begin{align*}
\braket{0|\rho_{\text{S}}|9},\braket{1|\rho_{\text{S}}|0},\braket{2|\rho_{\text{S}}|1},\braket{3|\rho_{\text{S}}|2},\braket{4|\rho_{\text{S}}|3},\braket{5|\rho_{\text{S}}|4},\braket{6|\rho_{\text{S}}|5},\braket{7|\rho_{\text{S}}|6},\braket{8|\rho_{\text{S}}|7},\braket{9|\rho_{\text{S}}|8}.
\end{align*}
Refer to Fig.~\ref{FIG5}, where the elements are shown in green within outlined regions.
\item If $U_{\text{S}}=X^2_{\text{S}}$, we have
\begin{align*}
\braket{0|\rho_{\text{S}}|8},\braket{1|\rho_{\text{S}}|9},\braket{2|\rho_{\text{S}}|0},\braket{3|\rho_{\text{S}}|1},\braket{4|\rho_{\text{S}}|2},\braket{5|\rho_{\text{S}}|3},\braket{6|\rho_{\text{S}}|4},\braket{7|\rho_{\text{S}}|5},\braket{8|\rho_{\text{S}}|6},,\braket{9|\rho_{\text{S}}|7}.
\end{align*}
Refer to Fig.~\ref{FIG5}, where the elements are shown in blue within outlined regions.
\item If $U_{\text{S}}=X^3_{\text{S}}$, we have
\begin{align*}
\braket{0|\rho_{\text{S}}|7},\braket{1|\rho_{\text{S}}|8},\braket{2|\rho_{\text{S}}|9},\braket{3|\rho_{\text{S}}|0},\braket{4|\rho_{\text{S}}|1},\braket{5|\rho_{\text{S}}|2},\braket{6|\rho_{\text{S}}|3},\braket{7|\rho_{\text{S}}|4},\braket{8|\rho_{\text{S}}|5},\braket{9|\rho_{\text{S}}|6}.
\end{align*}
Refer to Fig.~\ref{FIG5}, where the elements are shown in pink within outlined regions.
\item If $U_{\text{S}}=X^4_{\text{S}}$, we have
\begin{align*}
\braket{0|\rho_{\text{S}}|6},\braket{1|\rho_{\text{S}}|7},\braket{2|\rho_{\text{S}}|8},\braket{3|\rho_{\text{S}}|9},\braket{4|\rho_{\text{S}}|0},\braket{5|\rho_{\text{S}}|1},\braket{6|\rho_{\text{S}}|2},\braket{7|\rho_{\text{S}}|3},\braket{8|\rho_{\text{S}}|4},\braket{9|\rho_{\text{S}}|5}.
\end{align*}
Refer to Fig.~\ref{FIG5}, where the elements are shown in grey within outlined regions.
\item If $U_{\text{S}}=X^5_{\text{S}}$, we have
\begin{align*}
\braket{0|\rho_{\text{S}}|5},\braket{1|\rho_{\text{S}}|6},\braket{2|\rho_{\text{S}}|7},\braket{3|\rho_{\text{S}}|8},\braket{4|\rho_{\text{S}}|9},\braket{5|\rho_{\text{S}}|0},\braket{6|\rho_{\text{S}}|1},\braket{7|\rho_{\text{S}}|2},\braket{8|\rho_{\text{S}}|3},\braket{9|\rho_{\text{S}}|4}.
\end{align*}
Refer to Fig.~\ref{FIG5}, where the elements are shown in yellow within outlined regions.
\end{itemize}
Thus we need only six Pauli X-gates: $X^0_{\text{S}}$, $X^1_{\text{S}}$, $X^2_{\text{S}}$, $X^3_{\text{S}}$, $X^4_{\text{S}}$, and $X^5_{\text{S}}$  to fully characterize the unknown density matrix of a \emph{ten-dimensional} quantum system.\par


\section{S7. Precision in Density Matrix Characterization: Our Method vs. Existing Approaches}
Existing approaches in the literature employ \( d \) unitary evolution operators to reconstruct the full density matrix by accessing all \( d \) columns (or row). It is important to note that, in these methods, the Hermiticity of the density matrix does not reduce the number of required measurements. In contrast, our approach does not require full reconstruction of the density matrix; see S6 for detail discussion. To fully reconstruct the density matrix, we use Hermiticity condition of the density matrix. Accordingly, we first define the average estimation error for the existing methods, and then for our approach.\par
We assume that each matrix element $\braket{i|\rho_{\text{S}}|j}$ is a function of probabilities. Now we define the error in estimating the matrix element $\braket{i|\rho_{\text{S}}|j}$ as
\begin{align}
|\delta\braket{i|\rho_{\text{S}}|j}|^2=\delta \text{Re}[\braket{i|\rho_{\text{S}}|j}]^2+\delta \text{Im}[\braket{i|\rho_{\text{S}}|j}]^2,\tag{S7-1}\label{S7-1}
\end{align}
where 
\begin{equation}\tag{S7-2}\label{S7-2}
\begin{aligned}
\delta \text{Re}[\braket{i|\rho_{\text{S}}|j}]^2&=\sum_{l\in \mathcal{M}} \left(\frac{\partial \text{Re}[\braket{i|\rho_{\text{S}}|j}]}{\partial p(l_{\text{P}},i_{\text{S}})}\right)^2\delta^2p(l_{\text{P}},i_{\text{S}}),\\
\delta \text{Im}[\braket{i|\rho_{\text{S}}|j}]^2&=\sum_{l\in \mathcal{M}} \left(\frac{\partial \text{Im}[\braket{i|\rho_{\text{S}}|j}]}{\partial p(l_{\text{P}},i_{\text{S}})}\right)^2\delta^2p(l_{\text{P}},i_{\text{S}}),
\end{aligned}
\end{equation}
and $\mathcal{M}$ is the set of quantum pure states.
\subsubsection{Average error in estimating an unknown density matrix by existing approaches}
The error in estimating the unknown density matrix $\rho_{\text{S}}$ is defined as the sum of all the errors in estimating each matrix element $\braket{i|\rho_{\text{S}}|j}$ of $\rho_{\text{S}}$:
\begin{align}
\delta\rho_{\text{S}}=\sqrt{\sum_{i,j=0}^{d_{\text{S}}-1}|\delta\braket{i|\rho_{\text{S}}|j}|^2}.\tag{S7-3}\label{S7-3}
\end{align}

\subsubsection{Average error in estimating an unknown density matrix by our approach}
The error in estimating the unknown density matrix $\rho_{\text{S}}$ is defined as the sum of all the errors in estimating each matrix element of $\{\{\braket{i|\rho_{\text{S}}|(i-n)\,\,\,modulo\,\,\,d_{\text{S}}}\}_{i=0}^{d_{\text{S}}-1}\}_{n=0}^{F}$:
\begin{align}
\delta\rho_{\text{S}}=\sqrt{\sum_{n=0}^{F}\sum_{i=0}^{d_{\text{S}}-1}|\delta\braket{i|\rho_{\text{S}}|(i-n)\,\,\,modulo\,\,\,d_{\text{S}}}|^2},\tag{S7-4}\label{S7-4}
\end{align}
where $F=\frac{d_{\text{S}}}{2}$ or $\frac{d_{\text{S}}-1}{2}$ if $d_{\text{S}}$ is even or odd, respectively.

\subsection{S7.1. By our method given in Eq. \eqref{DFCDM-1}}
In Eq. \eqref{DFCDM-1}, we consider the unitary evolution operator to be such that $U_{\text{S}}\ket{i}=X_{\text{S}}^n\ket{i}=\ket{(i+n)\,\,modulo\,\,d_{\text{S}}}$, and ${X_{\text{S}}^n}^{\dagger}\ket{i}=\ket{(i-n)\,\,modulo\,\,d_{\text{S}}}=\ket{j}$. Then the real and imaginary parts of the element $\braket{i|\rho_{\text{S}}|(i-n)\,\,\,modulo\,\,\,d_{\text{S}}}$ are:
\begin{equation}\tag{S7.1-1}\label{S7.1-1}
\begin{aligned}
\text{Re}[\braket{i|\rho_{\text{S}}|(i-n)\,\,\,modulo\,\,\,d_{\text{S}}}]&=p(+_{\text{P}},i_{\text{S}})-p(-_{\text{P}},i_{\text{S}}),\\
\text{Im}[\braket{i|\rho_{\text{S}}|(i-n)\,\,\,modulo\,\,\,d_{\text{S}}}]&=p(+i_{\text{P}},i_{\text{S}})-p(-i_{\text{P}},i_{\text{S}}),
\end{aligned}
\end{equation}
where $p(\pm_{\text{P}},i_{\text{S}})=Tr[(\ketbra{\pm_{\text{P}}}{\pm_{\text{P}}}\otimes\Pi^i_{\text{S}})\rho(t)]$, $p(\pm i_{\text{P}},i_{\text{S}})=Tr[(\ketbra{\pm i_{\text{P}}}{\pm i_{\text{P}}}\otimes\Pi^i_{\text{S}})\rho(t)]$, and we have taken $\ket{\chi_{\text{P}}}=\frac{1}{\sqrt{2}}(\ket{0_{\text{P}}}+\ket{1_{\text{P}}})\implies\mathcal{N}_{\text{PS}}=1$ in Eq. \eqref{DFCDM-1}. Note that $\{\ket{\pm_{\text{P}}}\}$ and $\{\ket{\pm i_{\text{P}}}\}$ are eigenstates of the Pauli operators $\sigma_{\text{P}}^x$ and $\sigma_{\text{P}}^y$, respectively.\par
We assume that $N$ particles are used in determining $\braket{i|\rho_{\text{S}}|(i-n)\,\,\,modulo\,\,\,d_{\text{S}}}$. Since both the sets $\{\ket{\pm_{\text{P}}}\}$ and $\{\ket{\pm i_{\text{P}}}\}$ are measured, we allocate $\frac{N}{2}$ particles for the measurement of $\{\ket{\pm_{\text{P}}}\}$ and the remaining $\frac{N}{2}$ particles for the measurement of $\{\ket{\pm i_{\text{P}}}\}$. Also let $n_{l_{\text{P}},i_{\text{S}}}$ is the number of particles that have the post-measurement state $\ket{l_{\text{P}}}\in\mathcal{M}=\{\ket{\pm_{\text{P}}},\ket{\pm i_{\text{P}}}\}$ when the system is in $\ket{i_{\text{S}}}$ such that the probability is given by
\begin{align}
p(l_{\text{P}},i_{\text{S}})=\frac{n_{l_{\text{P}},i_{\text{S}}}}{N/2}=\frac{2}{N}n_{l_{\text{P}},i_{\text{S}}}.\tag{S7.1-2}\label{S7.1-2}
\end{align}
Now the variance of the measured probability $p(l_{\text{P}},i_{\text{S}})$ is given by
\begin{align}
\delta^2p(l_{\text{P}},i_{\text{S}})=\frac{4}{N^2}\delta^2n_{l_{\text{P}},i_{\text{S}}}=\frac{4}{N^2}n_{l_{\text{P}},i_{\text{S}}}=\frac{2}{N}p(l_{\text{P}},i_{\text{S}}),\tag{S7.1-3}\label{S7.1-3}
\end{align}
where we considered that the statistic follows the Poissonian statistic and hence the variance of $n_{l_{\text{P}},i_{\text{S}}}$ is equal to $n_{l_{\text{P}},i_{\text{S}}}$, and used Eq. \eqref{S7.1-2}. The error in estimating the matrix element $\braket{i|\rho_{\text{S}}|(i-n)\,\,\,modulo\,\,\,d_{\text{S}}}$ defined in Eq. \eqref{S7-1} using Eqs. \eqref{S7.1-1} and \eqref{S7.1-3} is given by
\begin{align}
|\delta\braket{i|\rho_{\text{S}}|(i-n)\,\,\,modulo\,\,\,d_{\text{S}}}|^2&=\frac{2}{N}\left[\{p(+_{\text{P}},i_{\text{S}})+p(-_{\text{P}},i_{\text{S}})\}+\{p(+i_{\text{P}},i_{\text{S}})+p(-i_{\text{P}},i_{\text{S}})\}\right]\nonumber\\
&=\frac{2}{N}\left[\braket{i|\rho_{\text{S}}|i}+\braket{i|X^n_{\text{S}}\rho_{\text{S}}{X^n_{\text{S}}}^{\dagger}|i}\right]\nonumber\\
&=\frac{2}{N}\left[\braket{i|\rho_{\text{S}}|i}+\braket{(i-n)\,\,\,modulo\,\,\,d_{\text{S}}|\rho_{\text{S}}|(i-n)\,\,\,modulo\,\,\,d_{\text{S}}}\right],\tag{S7.1-4}\label{S7.1-4}
\end{align}
where we have substituted the values of $p(\pm_{\text{P}},i_{\text{S}})$ and $p(\pm i_{\text{P}},i_{\text{S}})$ defined in Eq. \eqref{S7.1-1}, and $U_{\text{S}}$ has been taken to be such that $U_{\text{S}}\ket{i}=X_{\text{S}}^n\ket{i}=\ket{(i+n)\,\,modulo\,\,d_{\text{S}}}$, and ${X_{\text{S}}^n}^{\dagger}\ket{i}=\ket{(i-n)\,\,modulo\,\,d_{\text{S}}}$.\par
Now the  error in estimating $\rho_{\text{S}}$ as given in Eq.~\eqref{S7-4} is:
\begin{align}
\delta\rho_{\text{S}}&=\sqrt{\sum_{n=0}^{F}\sum_{i=0}^{d_{\text{S}}-1}|\delta\braket{i|\rho_{\text{S}}|(i-n)\,\,\,modulo\,\,\,d_{\text{S}}}|^2}\nonumber\\
&=\sqrt{\sum_{n=0}^{F}\sum_{i=0}^{d_{\text{S}}-1}\frac{2}{N}\left[\braket{i|\rho_{\text{S}}|i}+\braket{(i-n)\,\,\,modulo\,\,\,d_{\text{S}}|\rho_{\text{S}}|(i-n)\,\,\,modulo\,\,\,d_{\text{S}}}\right]}\nonumber\\
&=2\sqrt{\frac{F+1}{N}}=\begin{cases}
    2\sqrt{\frac{d_{\text{S}}+2}{2N}},\quad\quad\quad if\,\,\,d_{\text{S}}\,\,\,is\,\,\,even,\\
    2\sqrt{\frac{d_{\text{S}}+1}{2N}},\quad\quad\quad if\,\,\,d_{\text{S}}\,\,\,is\,\,\,odd.
    \end{cases}\tag{S7.1-6}\label{S7.1-6}
\end{align}

\subsection{S7.2. By the method given in Ref. \cite{Vallone-2018}}
In this work \cite{Vallone-2018}, the authors have considered a system ($S$) and two qubit probes $A$ and $B$. Then the real and imaginary parts of the element $\braket{i|\rho_{\text{S}}|j}$ are given by:
\begin{equation}\tag{S7.2-1}\label{S7.2-1}
\begin{aligned}
\braket{i|\rho_{\text{S}}|i}&=16\mathcal{N}_{AB}^2Tr\left[\Pi_{\text{S}}^{j}\otimes\Pi^1_A\otimes\Pi^1_B\left(U_BU_A^i\left\{\rho_{\text{S}}\otimes\ketbra{0_A}{0_A}\otimes\ketbra{0_B}{0_B}\right\}{U_A^i}^{\dagger}U_B^{\dagger}\right)\right],\quad\quad\forall i\\
\text{Re}[\braket{i|\rho_{\text{S}}|j}]&=-2\mathcal{N}_{AB}Tr\left[\Pi_{\text{S}}^{j}\otimes Y_A\otimes Y_B\left(U_BU_A^i\left\{\rho_{\text{S}}\otimes\ketbra{0_A}{0_A}\otimes\ketbra{0_B}{0_B}\right\}{U_A^i}^{\dagger}U_B^{\dagger}\right)\right],\quad\quad i\neq j\\
\text{Im}[\braket{i|\rho_{\text{S}}|j}]&=2\mathcal{N}_{AB}Tr\left[\Pi_{\text{S}}^{j}\otimes X_A\otimes Y_B\left(U_BU_A^i\left\{\rho_{\text{S}}\otimes\ketbra{0_A}{0_A}\otimes\ketbra{0_B}{0_B}\right\}{U_A^i}^{\dagger}U_B^{\dagger}\right)\right], \quad\quad\,\,\, i\neq j
\end{aligned}
\end{equation}
where $\Pi^j_{\text{S}}=\ketbra{j_{\text{S}}}{j_{\text{S}}}$, $\Pi^1_A=\ketbra{1_A}{1_A}$, $\Pi^1_B=\ketbra{1_B}{1_B}$, and $X_A$,  $Y_A$, and  $Y_B$ are the Pauli operators. Here $\mathcal{N}_{AB}=d_{\text{S}}/(4\sin^2\theta)$, $U_A^i=e^{-i\theta\Pi^i_{\text{S}}\otimes Y_A}\otimes I_B$, and $U_B=e^{-i\theta\Pi^{b_0}_{\text{S}}\otimes Y_B}\otimes I_A$, where $\Pi_{\text{S}}^{b^0}=\ketbra{b^0_{\text{S}}}{b^0_{\text{S}}}$, and $\ket{b^0_{\text{S}}}=\frac{1}{\sqrt{d_{\text{S}}}}\sum_{i=0}^{d_{\text{S}}-1}\ket{i}$. \par
After substituting the spectral forms of the Pauli operators,  Eq. \eqref{S7.2-1} becomes
\begin{equation}\tag{S7.2-2}\label{S7.2-2}
\begin{aligned}
\braket{i|\rho_{\text{S}}|i}&=16\mathcal{N}_{AB}^2 p(j_{\text{S}},1_A,1_B),\quad\quad\quad\quad\quad\quad\quad\quad\quad\quad\quad\quad\quad\quad\quad\quad\quad\quad\quad\quad\quad\quad\quad\quad\quad\quad\quad \forall i\\
\text{Re}[\braket{i|\rho_{\text{S}}|j}]&=-2\mathcal{N}_{AB}[p(j_{\text{S}},+i_A,+i_B)-p(j_{\text{S}},+i_A,-i_B)-p(j_{\text{S}},-i_A,+i_B)+p(j_{\text{S}},-i_A,-i_B)],\quad\quad i\neq j\\
\text{Im}[\braket{i|\rho_{\text{S}}|j}]&=2\mathcal{N}_{AB}[p(j_{\text{S}},+_A,+i_B)-p(j_{\text{S}},+_A,-i_B)-p(j_{\text{S}},-_A,+i_B)+p(j_{\text{S}},-_A,-i_B)],\quad\quad\quad\quad i\neq j
\end{aligned}
\end{equation}
where $p(j_{\text{S}},l_A,l_B)=Tr\left[\Pi_{\text{S}}^{j}\otimes\Pi^l_A\otimes\Pi^l_B\left(U_BU_A^i\left\{\rho_{\text{S}}\otimes\ketbra{0_A}{0_A}\otimes\ketbra{0_B}{0_B}\right\}{U_A^i}^{\dagger}U_B^{\dagger}\right)\right]$ is the probability that the system and the two qubit probes will be in the state $\ket{j_{\text{S}}}\otimes\ket{l_A}\otimes\ket{l_B}$, where $\ket{l_A}\in\mathcal{M}_A=\{\ket{0_A},\ket{1_A},\ket{\pm_A},\ket{\pm i_A}\}$, $\ket{l_B}\in\mathcal{M}_B=\{\ket{0_B},\ket{1_B},\ket{\pm_B},\ket{\pm i_B}\}$ such that
\begin{align}
\sum_{j=0}^{d_{\text{S}}-1}\sum_{l_A=\{\ket{\pm_A}\}}\sum_{l_B=\{\ket{i\pm_B}\}}p(j_{\text{S}},l_A,l_B)=1,\quad\quad \sum_{j=0}^{d_{\text{S}}-1}\sum_{l_A=\{\ket{i\pm_A}\}}\sum_{l_B=\{\ket{i\pm_B}\}}p(j_{\text{S}},l_A,l_B)=1.
\end{align}
We assume that $N$ particles are used in determining $\braket{i|\rho_{\text{S}}|j}$, also assuming that one single particle acts as the system as well as two qubit probes. Since three sets $\{\ket{0_A},\ket{1_A}\}$, $\{\ket{\pm_A}\}$ and $\{\ket{\pm i_A}\}$ are measured, we allocate $\frac{N}{3}$ particles for the measurement of $\{\ket{0_A},\ket{1_A}\}$, $\frac{N}{3}$ particles for the measurement of $\{\ket{\pm_A}\}$ and the remaining $\frac{N}{3}$ particles for the measurement of $\{\ket{\pm i_A}\}$. Let $n_{j_{\text{S}},l_A,l_B}$ is the number of particles that have the post-measurement state $\ket{l_A}\otimes\ket{l_B}\in\mathcal{M}_A\otimes\mathcal{M}_B$ when the system is in $\ket{j_{\text{S}}}$ such that the probability is given by
\begin{align}
p(j_{\text{S}},l_A,l_B)=\frac{n_{j_{\text{S}},l_A,l_B}}{N/3}=\frac{3}{N}n_{j_{\text{S}},l_A,l_B}.\tag{S7.2-3}\label{S7.2-3}
\end{align}
Now the variance of the measured probability $p(j_{\text{S}},l_A,l_B)$ is given by
\begin{align}
\delta^2p(j_{\text{S}},l_A,l_B)=\frac{9}{N^2}\delta^2n_{j_{\text{S}},l_A,l_B}=\frac{9}{N^2}n_{j_{\text{S}},l_A,l_B}=\frac{3}{N}p(j_{\text{S}},l_A,l_B),\tag{S7.2-4}\label{S7.2-4}
\end{align}
where we considered that the  statistic follows the Poissonian statistic and hence the variance of $n_{j_{\text{S}},l_A,l_B}$ is equal to $n_{j_{\text{S}},l_A,l_B}$, and used Eq. \eqref{S7.2-3}. The error in estimating the matrix element $\braket{i|\rho_{\text{S}}|j}$ defined in Eq. \eqref{S7-1} (instead of bipartite probabilities, consider tripartite probabilities in this case) using Eqs. \eqref{S7.2-2} and \eqref{S7.2-4} is given by
\begin{equation}\tag{S7.2-5}\label{S7.2-5}
\begin{aligned}
|\delta\braket{i|\rho_{\text{S}}|i}|^2&=\frac{3\times16^2}{N}\mathcal{N}_{AB}^4p(j_{\text{S}},1_A,1_B)=\frac{48}{N}\mathcal{N}_{AB}^2\braket{i|\rho_{\text{S}}|i}, \quad\quad\quad\quad\quad\quad\quad\quad\quad\quad\quad\forall i\\
|\delta\braket{i|\rho_{\text{S}}|j}|^2&=\frac{12}{N}\mathcal{N}_{AB}^2\sum_{l_A\in\{\ket{\pm_A},\ket{\pm i_A}\}}\sum_{l_B\in\{\ket{\pm i_B}\}}p(j_{\text{S}},l_A,l_B)=\frac{24}{N}\mathcal{N}_{AB}^2p(j_{\text{S}}),\quad\quad\quad\quad i\neq j
\end{aligned}
\end{equation}
where $\sum_{l_A\in\{\ket{\pm_A}\}}\sum_{l_B\in\{\ket{\pm i_B}\}}p(j_{\text{S}},l_A,l_B)=p(j_{\text{S}})=\sum_{l_A\in\{\ket{\pm i_A}\}}\sum_{l_B\in\{\ket{\pm i_B}\}}p(j_{\text{S}},l_A,l_B)$ is the marginal probability.\par 
Now the  error in estimating $\rho_{\text{S}}$ as given in Eq.~\eqref{S7-3} is:
\begin{align}
\delta\rho_{\text{S}}&=\sqrt{\sum_{i,j=0}^{d_{\text{S}}-1}|\delta\braket{i|\rho_{\text{S}}|j}|^2}\nonumber\\
&=\sqrt{\frac{48}{N}\mathcal{N}_{AB}^2\sum_{j=0}^{d_{\text{S}}-1}\braket{j|\rho_{\text{S}}|j}+\frac{24}{N}\mathcal{N}_{AB}^2\sum_{i\neq j}^{d_{\text{S}}-1}p(j_{\text{S}})}\nonumber\\
&=\sqrt{\frac{48}{N}\mathcal{N}_{AB}^2\sum_{j=0}^{d_{\text{S}}-1}\braket{j|\rho_{\text{S}}|j}+\frac{24}{N}\mathcal{N}_{AB}^2\left[\sum_{i,j=0}^{d_{\text{S}}-1}p(j_{\text{S}})-\sum_{j=0}^{d_{\text{S}}-1}p(j_{\text{S}})\right]}\nonumber\\
&=\frac{\sqrt{6}d_{\text{S}}}{2\sin^2\theta}\sqrt{\frac{d_{\text{S}}+1}{N}}.\tag{S7.2-6}\label{S7.2-6}
\end{align}
Clearly, the error in estimating $\rho_{\text{S}}$ using this method is greater than that obtained with our approach in Eq.~\eqref{S7.1-6} for any value of $\theta$. When $\theta$ approaches zero, the error $\delta\rho_{\text{S}}$ becomes substantially high. Furthermore, if the second probe is an additional particle, the total number of particles used is doubled, making the estimation of $\rho_{\text{S}}$ via the method in Eq.~\eqref{S7.2-6} even more error-prone, whereas our method requires only a single probe. There are some other disadvantages of their method discussed in the main text.

\subsection{S7.3. By the method given in Ref. \cite{Xu-Zhou-2024}}
In this work \cite{Xu-Zhou-2024}, the authors have considered a qubit probe and a system ($S$). Then the real and imaginary parts of the element $\rho_{ij}$ are given by:
The real and imaginary parts of the element $\braket{i|\rho_{\text{S}}|j}$ are:
\begin{equation}\tag{S7.3-1}\label{S7.3-1}
\begin{aligned}
\text{Re}[\braket{i|\rho_{\text{S}}|j}]&=\frac{1}{2\sin 2\theta}[p(+_{\text{P}},i_{\text{S}})-p(-_{\text{P}},i_{\text{S}})+p(+_{\text{P}},j_{\text{S}})-p(-_{\text{P}},j_{\text{S}})],\\
\text{Im}[\braket{i|\rho_{\text{S}}|j}]&=\frac{1}{2\sin 2\theta}[-p(+i_{\text{P}},i_{\text{S}})-p(-i_{\text{P}},i_{\text{S}})+p(+i_{\text{P}},j_{\text{S}})-p(-i_{\text{P}},j_{\text{S}})],
\end{aligned}
\end{equation}
where $p(\pm_{\text{P}},i_{\text{S}})=Tr[(\ketbra{\pm_{\text{P}}}{\pm_{\text{P}}}\otimes\Pi^i_{\text{S}})\rho(t)]$, $p(\pm i_{\text{P}},i_{\text{S}})=Tr[(\ketbra{\pm i_{\text{P}}}{\pm i_{\text{P}}}\otimes\Pi^i_{\text{S}})\rho(t)]$.\par
We assume that $N$ particles are used in determining $\braket{i|\rho_{\text{S}}|j}$. Since both the sets $\{\ket{\pm_{\text{P}}}\}$ and $\{\ket{\pm i_{\text{P}}}\}$ are measured for $i_{\text{S}}$ and $j_{\text{S}}$ indices, we allocate:
\begin{itemize}
\item $\frac{N}{4}$ particles for the measurement of $\{\ket{\pm_{\text{P}}}\}$ for a given $i_{\text{S}}$,
\item $\frac{N}{4}$ particles for the measurement of $\{\ket{\pm i_{\text{P}}}\}$ for a given $i_{\text{S}}$,
\item $\frac{N}{4}$ particles for the measurement of $\{\ket{\pm_{\text{P}}}\}$ for a given $j_{\text{S}}$,
\item and the remaining $\frac{N}{4}$ particles for the measurement of $\{\ket{\pm i_{\text{P}}}\}$ for a given $j_{\text{S}}$.
\end{itemize}
Also let $n_{l_{\text{P}},i_{\text{S}}}$ is the number of particles that have the post-measurement state $\ket{l_{\text{P}}}\in\mathcal{M}=\{\ket{\pm_{\text{P}}},\ket{\pm i_{\text{P}}}\}$ when the system is in $\ket{i_{\text{S}}}$ such that the probability is given by
\begin{align}
p(l_{\text{P}},i_{\text{S}})=\frac{n_{l_{\text{P}},i_{\text{S}}}}{N/4}=\frac{4}{N}n_{l_{\text{P}},i_{\text{S}}}.\tag{S7.3-2}\label{S7.3-2}
\end{align}
Now the variance of the measured probability $p(l_{\text{P}},i_{\text{S}})$ is given by
\begin{align}
\delta^2p(l_{\text{P}},i_{\text{S}})=\frac{16}{N^2}\delta^2n_{l_{\text{P}},i_{\text{S}}}=\frac{16}{N^2}n_{l_{\text{P}},i_{\text{S}}}=\frac{4}{N}p(l_{\text{P}},i_{\text{S}}),\tag{S7.3-3}\label{S7.3-3}
\end{align}
where we considered that the statistic follows the Poissonian statistic and hence the variance of $n_{l_{\text{P}},i_{\text{S}}}$ is equal to $n_{l_{\text{P}},i_{\text{S}}}$, and used Eq. \eqref{S7.3-2}. Similarly, one can calculate the variance of the measured probability $p(l_{\text{P}},j_{\text{S}})$ also. The error in estimating the matrix element $\braket{i|\rho_{\text{S}}|j}$ defined in Eq. \eqref{S7-1} using Eqs. \eqref{S7.3-1} and \eqref{S7.3-3} is given by
\begin{align}
|\delta\braket{i|\rho_{\text{S}}|j}|^2&=\frac{1}{4\sin^22\theta}\frac{4}{N}\left[\sum_{l_{\text{P}}\in\{\ket{\pm_{\text{P}}}\}}\left\{p(l_{\text{P}},i_{\text{S}})+p(l_{\text{P}},j_{\text{S}})\right\}+\sum_{l_{\text{P}}\in\{\ket{\pm i_{\text{P}}}\}}\left\{p(l_{\text{P}},i_{\text{S}})+p(l_{\text{P}},j_{\text{S}})\right\}\right]\nonumber\\
&=\frac{1}{\sin^22\theta}\frac{2}{N}\left[p(i_{\text{S}})+p(j_{\text{S}})\right],\tag{S7.3-4}\label{S7.3-4}
\end{align}
where $\sum_{l_{\text{P}}\in\{\ket{\pm_{\text{P}}}\}}p(l_{\text{P}},i_{\text{S}})=p(i_{\text{S}})=\sum_{l_{\text{P}}\in\{\ket{\pm i_{\text{P}}}\}}p(l_{\text{P}},i_{\text{S}})$ is the marginal probability. Similarly for $j_{\text{S}}$ index also. \par
Now the  error in estimating $\rho_{\text{S}}$ as given in Eq.~\eqref{S7-3} is:
\begin{align}
\delta\rho_{\text{S}}&=\sqrt{\sum_{i,j=0}^{d_{\text{S}}-1}|\delta\braket{i|\rho_{\text{S}}|j}|^2}\nonumber\\
&=\sqrt{\frac{1}{\sin^22\theta}\frac{2}{N}\sum_{i,j=0}^{d_{\text{S}}-1}\left[p(i_{\text{S}})+p(j_{\text{S}})\right]}\nonumber\\
&=\frac{2}{\sin2\theta}\sqrt{\frac{d_{\text{S}}}{N}}.\tag{S7.3-5}\label{S7.3-5}
\end{align}
When \(\theta \to 0\), the error \(\delta \rho_{\text{S}}\) becomes significantly larger than that of our approach in Eq.~\eqref{S7.1-6}. For \(\theta = \frac{\pi}{4}\) and $d_{\text{S}}=2$, their result matches exactly with ours. Several important observations can be made regarding the method of Ref.~\cite{Xu-Zhou-2024}:  
($i$) \(\theta = \frac{\pi}{4}\) does not correspond to a strong measurement; the strong measurement limit is achieved at \(\theta = \frac{\pi}{2}\). Therefore, to attain optimal precision in their method, one must set \(\theta = \frac{\pi}{4}\), thus compromising the strength of the strong interaction.  
($ii$) A Hadamard transformation is required for each system projector \(\Pi_{\text{S}}^{i} = \ketbra{i}{i}\).  
($iii$) The method requires \(d_{\text{S}}\) distinct unitary operations based on the projectors \(\{\Pi_{\text{S}}^{i}\}\).  
Our framework avoids such complexities; see the main text for a detailed discussion.

\section{S8. Improving the accuracy of weak value calculation}
\subsection{A better approximation for determining weak values}
To obtain Eq.~(\ref{WV-2}), second and higher-order terms were discarded. In the following, we retain terms up to second order, discarding only third and higher-order contributions. We begin by expanding Eq.~(\ref{WV-1}) without approximation:
\begin{align}
\braket{\phi_{\text{S}}|\widetilde{U}_{\text{S}}|\psi_{\text{S}}} &= \braket{\phi_{\text{S}}|e^{-i(\theta_1 - \theta_2)A}|\psi_{\text{S}}} \nonumber\\
&=\braket{\phi_{\text{S}}|\psi_{\text{S}}} \left[ 1 - i\delta\theta\frac{\braket{\phi_{\text{S}}|A|\psi_{\text{S}}}}{\braket{\phi_{\text{S}}|\psi_{\text{S}}}}-\frac{1}{2!}\delta\theta^2\frac{\braket{\phi_{\text{S}}|A^2|\psi_{\text{S}}}}{\braket{\phi_{\text{S}}|\psi_{\text{S}}}}+\cdots\right], \tag{S8-1}\label{S8-1}
\end{align}
where $\delta\theta=\theta_1-\theta_2$. Similarly, 
\begin{align}
\braket{\phi_{\text{S}}|\widetilde{U}^{\dagger}_{\text{S}}|\psi_{\text{S}}} &= \braket{\phi_{\text{S}}|e^{i(\theta_1 - \theta_2)A}|\psi_{\text{S}}} \nonumber\\
&=\braket{\phi_{\text{S}}|\psi_{\text{S}}} \left[ 1 + i\delta\theta\frac{\braket{\phi_{\text{S}}|A|\psi_{\text{S}}}}{\braket{\phi_{\text{S}}|\psi_{\text{S}}}}-\frac{1}{2!}\delta\theta^2\frac{\braket{\phi_{\text{S}}|A^2|\psi_{\text{S}}}}{\braket{\phi_{\text{S}}|\psi_{\text{S}}}}+\cdots\right]. \tag{S8-2}\label{S8-2}
\end{align}
Now by subtracting Eq. (\ref{S8-1}) from Eq. (\ref{S8-2}), we have
\begin{align}
\braket{\phi_{\text{S}}|\widetilde{U}^{\dagger}_{\text{S}}|\psi_{\text{S}}}-\braket{\phi_{\text{S}}|\widetilde{U}_{\text{S}}|\psi_{\text{S}}}&=2\braket{\phi_{\text{S}}|\psi_{\text{S}}} \left[i\delta\theta\frac{\braket{\phi_{\text{S}}|A|\psi_{\text{S}}}}{\braket{\phi_{\text{S}}|\psi_{\text{S}}}}+\mathcal{O}(\delta\theta^3)\right]\nonumber\\
&\approx 2\braket{\phi_{\text{S}}|\psi_{\text{S}}}i\delta\theta\frac{\braket{\phi_{\text{S}}|A|\psi_{\text{S}}}}{\braket{\phi_{\text{S}}|\psi_{\text{S}}}}, \tag{S8-3}\label{S8-3}
\end{align}
where the third and higher order terms are discarded, and thus we obtain the weak value:
\begin{align}
\frac{\braket{\phi_{\text{S}}|A|\psi_{\text{S}}}}{\braket{\phi_{\text{S}}|\psi_{\text{S}}}}=\frac{1}{2i\delta\theta}\left[\frac{\braket{\phi_{\text{S}}|\widetilde{U}^{\dagger}_{\text{S}}|\psi_{\text{S}}}}{\braket{\phi_{\text{S}}|\psi_{\text{S}}}}-\frac{\braket{\phi_{\text{S}}|\widetilde{U}_{\text{S}}|\psi_{\text{S}}}}{\braket{\phi_{\text{S}}|\psi_{\text{S}}}}\right].\tag{S8-4}\label{S8-4}
\end{align}
By substituting the weak values of the unitary operators $\widetilde{U}^{\dagger}_{\text{S}}$ and $\widetilde{U}_{\text{S}}$ from Eq.~(\ref{DFCUO-1}) with $U_{\text{S}}=I_{\text{S}}$ into Eq.~(\ref{S8-4}), a more accurate estimate of the weak value of the observable $A$ is obtained.
\subsection{Method for exact determination of weak values}
We first assume that the observable $A$ has $d_{\text{S}}$ non-degenerate, known eigenvalues. Under this condition, the Lagrange interpolation method allows us to express the exponential of $A$ in a compact form~\cite{Moler-1978}:
\begin{align*}
e^{-i\theta A} = \sum_{k=1}^{d_{\text{S}}} e^{-i\lambda_k \theta} \prod_{\substack{l=1\\l \neq k}}^{d_{\text{S}}} \frac{A - \lambda_l I}{\lambda_k - \lambda_l}.
\end{align*}
It can be seen that, for a $d_{\text{S}}$-dimensional Hilbert space, the modular value of an observable $A$ defined in Eq. (\ref{MV-1}) can be expressed in terms of weak values of $A$ up to order $d_{\text{S}} - 1$ as:
\begin{align}
\braket{A_m}_{\psi}^{\phi} = \Lambda + \Lambda'\braket{A_w}_{\psi}^{\phi} + \Lambda''\braket{A_w^2}_{\psi}^{\phi} + \cdots + \Lambda^{(d_{\text{S}}-1)\prime} \braket{A_w^{d_{\text{S}}-1}}_{\psi}^{\phi}. \tag{S8-5}\label{S8-5}
\end{align}
Here, the coefficients $\Lambda, \Lambda', \Lambda'', \dots, \Lambda^{(d_{\text{S}}-1)\prime}$ depend on the eigenvalues of $A$ and a parameter $\theta$. By solving the resulting matrix equation from Eq.~(\ref{S8-5}) for suitably chosen values of $\theta$, one can extract not only the weak value $\braket{A_w}_{\psi}^{\phi}$, but also its higher moments $\braket{A_w^n}_{\psi}^{\phi}$ for $n = 2, 3, \dots, d_{\text{S}} - 1$. Higher-moment weak values encode significant information about quantum systems, including weak probabilities, which can be employed to address the \emph{quantum three-box problem}~\cite{Aharonov-2002,Resch-Lundeen-Steinberg-2024}.\par
As an illustrative example, we consider a spin-1 system with $d_{\text{S}} = 3$, where the observable $A$ has eigenvalues $\{-1, 0, 1\}$. In this case, the exponential $e^{-iA\theta}$ simplifies to the following operator polynomial \cite{Ho-Imoto-2016-ModularFinite}:
\begin{align}
e^{-i\theta A} = I - i \sin(\theta) A + [\cos(\theta) - 1] A^2. \tag{S8-6}\label{S8-6}
\end{align}
By substituting Eq.~(\ref{S8-6}) into the expression of modular value defined in Eq.~(\ref{MV-1}), the modular value of $A$ becomes
\begin{align}
\braket{A_m}_{\psi}^{\phi} = 1 - i \sin(\theta) \braket{A_w}_{\psi}^{\phi} + [\cos(\theta) - 1] \braket{A_w^2}_{\psi}^{\phi}, \tag{S8-7}\label{S8-7}
\end{align}
where $\Lambda = 1$, $\Lambda' = -i \sin(t)$, and $\Lambda'' = \cos(t) - 1$ are the coefficients corresponding to the $I$, $A$, and $A^2$, respectively. As Eq.~(\ref{S8-7}) 
contains only two unknown parameters $\braket{A_w}_{\psi}^{\phi}$ and $\braket{A_w^2}_{\psi}^{\phi}$, we need two different values of $\theta$ to solve the Eq.~(\ref{S8-7}). By doing so, one obtains the weak value $\braket{A_w}_{\psi}^{\phi}$ without any approximations. 
\end{document}